\documentclass[journal]{IEEEtran}
\IEEEoverridecommandlockouts

\usepackage{cite}
\usepackage{amsmath,amssymb,amsfonts}
\usepackage{algorithm}
\usepackage{algorithmic}
\usepackage{diagbox}
\usepackage{subcaption}
\usepackage{adjustbox}
\usepackage{colortbl}

\usepackage[normalem]{ulem}
\usepackage{xcolor}

\usepackage[normalem]{ulem}

\usepackage{graphicx} 
\usepackage{subcaption}

\usepackage{stfloats}  
\usepackage{caption}   
\usepackage{footmisc}

\usepackage{multirow} 
\usepackage{pifont}
\usepackage{changes}
\definechangesauthor[name={Tongyu Song}, color=orange]{AF}

\usepackage{makecell}
\usepackage{graphicx}
\usepackage{hyperref}
\usepackage{textcomp}
\usepackage{xcolor}

\usepackage{microtype}

\def\BibTeX{{\rm B\kern-.05em{\sc i\kern-.025em b}\kern-.08em
    T\kern-.1667em\lower.7ex\hbox{E}\kern-.125emX}}
\begin{document}

\title{Two-Layer Reinforcement Learning-Assisted Joint Beamforming and Trajectory Optimization for Multi-UAV Downlink Communications}
\author{Ruiqi Wang, 
        Esraa M. Ghourab,~\IEEEmembership{Senior Member,~IEEE,} 
        Omar Alhussein,~\IEEEmembership{Senior Member,~IEEE,} 
        Yuzhi Yang,~\IEEEmembership{Member,~IEEE,} 
        Jing Ren,~\IEEEmembership{Member,~IEEE,} 
        Shizhong Xu,~\IEEEmembership{Member,~IEEE,} 
        and Sami Muhaidat,~\IEEEmembership{Senior Member,~IEEE}
        
\thanks{Corresponding author: Esraa M. Ghourab.}
\thanks{Ruiqi Wang, Jing Ren and Shizhong Xu are with the School of Information and Communication Engineering, University of Electronic Science and Technology of China, Chengdu 611731, China. E-mail: 202421011308@std.uestc.edu.cn; renjing@uestc.edu.cn; xsz@uestc.edu.cn.}
\thanks{Esraa M. Ghourab and Omar Alhussein are with KU 6G Research Center, Department of Computer Science, Khalifa University, UAE. E-mail: essra.ghourab@ku.ac.ae, omar.alhussein@ku.ac.ae.}
\thanks{Yuzhi Yang is with the institute of digital future, Khalifa University, Abu Dhabi, UAE. E-mail: yuzhi.yang@ku.ac.ae.}
\thanks{Sami Muhaidat is with KU 6G Research Center, Department of Computer and Information Engineering, Khalifa University, UAE. E-mail: sami.muhaidat@ku.ac.ae.}}

\maketitle
\begin{abstract}
Unmanned aerial vehicles (UAVs) are pivotal for future 6G non-terrestrial networks; however, their high mobility creates a complex coupled optimization problem for beamforming and trajectory design. 
Existing numerical methods suffer from prohibitive latency, while standard deep learning often ignores dynamic interference topology, which limits their scalability. 
To address these issues, this paper proposes a hierarchically decoupled framework that integrates graph neural networks (GNNs) with multi-agent reinforcement learning. 
On the fast timescale, we formulate a time-varying heterogeneous graph to model UAV–user associations and intra-cluster interference, and develop a GraphNorm-enhanced GNN beamformer that explicitly learns interference coupling patterns, enabling low-latency inference suitable for dynamic channel environments.
On the slow timescale, trajectory planning is modeled as a decentralized partially observable Markov decision process and solved via the multi-agent proximal policy optimization algorithm under the centralized training with decentralized execution paradigm, allowing cooperative mobility control under partial observations.
Extensive simulation results demonstrate that the proposed framework significantly outperforms conventional optimization heuristics and deep learning baselines in terms of achievable sum rate, convergence behavior, and generalization
across various network settings.
\end{abstract}


\section{Introduction}
Unmanned aerial vehicles (UAVs) are key components of future 6G non-terrestrial networks, enabling flexible three-dimensional (3D) deployment for on-demand wireless coverage \cite{survey1}.
However, UAV spatiotemporal trajectories directly dictate channel topology, creating a trade-off between mobility constraints and communication performance.
Conversely, finite beamforming capabilities and transmit power limits restrict feasible flight trajectories during interference mitigation \cite{survey4}.
This bidirectional coupling between mobility control and communication resource allocation results in a highly non-convex, dynamic optimization problem, further complicated in multi-UAV scenarios by cooperative interference management and collision avoidance requirements.

To tackle this problem, existing studies primarily rely on iterative numerical optimization \cite{survey4}.
While providing structured solutions to the coupled design, their iterative nature and high-dimensional search space incur substantial computational overhead, hindering real-time deployment in dynamic environments.
Reinforcement learning (RL)-based approaches offer faster inference for such complex problems.
However, two fundamental challenges remain.
First, standard end-to-end architectures are typically structure-agnostic, struggling to capture the intrinsic graph topology of user association.
Second, an inherent timescale mismatch exists between millisecond-level small-scale fading channels and second-level UAV mechanical mobility.
Consequently, treating beamforming and trajectory control synchronously creates a high-dimensional hybrid action space difficult to optimize efficiently.
To address these challenges, we propose a hierarchically decoupled framework separating the joint optimization into a fast timescale inner-loop for instantaneous beamforming and a slow timescale outer-loop for long-term trajectory planning.

For the inner-loop beamforming, the multi-UAV environment features dynamic graph-structured topologies capturable by a graph neural network (GNN).
Unlike conventional neural networks with fixed-dimensional inputs, GNNs leverage permutation invariance and parameter sharing to explicitly capture dynamic and irregular interference topologies \cite{gnn5}\cite{gnn6}.
Building upon this property, we propose a topology-aware beamformer directly mapping instantaneous channel state information (CSI) to beamforming vectors.
Specifically, we model the dynamic UAV-user system as a time-varying heterogeneous graph, where nodes represent UAVs/users, and edges encode communication/interference relationships.
Leveraging a heterogeneous message passing mechanism, the proposed beamformer efficiently captures the coupled interference patterns.
%Furthermore, we integrate GraphNorm to stabilize feature distributions across varying local graph scales, thereby empowering the model with robust generalization capabilities across diverse network topologies.

For the outer-loop trajectory planning, the problem is a complex sequential decision-making process characterized by non-stationarity arising from the concurrent policy learning of multiple UAV agents.
Multi-agent reinforcement learning (MARL) under the centralized training with decentralized execution (CTDE) paradigm is well-suited for such cooperative control problems \cite{CTDE}.
Within this framework, we employ the multi-agent proximal policy optimization (MAPPO) algorithm for superior stability in stochastic policy optimization \cite{MAPPO}.
However, directly applying standard MAPPO to the multi-UAV mission scenario presents two key challenges.
First, sparse feedback is inherently tied to long-term trajectory planning. 
Second, the asynchronous nature of mission completions leads to training instability, as different UAVs reach their destinations at different time steps based on flight distances.
To address these challenges, we enhance the algorithm with a novel reachability-aware reward mechanism coupled with an arrival masking scheme.
This dual-mechanism provides dense, step-wise supervision on flight feasibility while effectively filtering out invalid gradients from completed agents, thereby complementing the global value estimation.
This combined design improves learning efficiency, enabling agents to acquire coordinated flight behaviors guided by performance feedback from the inner-loop beamformer.

The proposed topology-aware GNN beamformer and feasibility-enhanced MAPPO trajectory planner constitute a unified closed-loop framework.
Within this architecture, the inner-loop beamformer evaluates instantaneous communication performance given the current network topology, providing reward signals to outer-loop agents.
This enables the trajectory policy to adaptively balance mobility constraints, collision avoidance, and long-term communication performance, facilitating coordinated decision-making.

The main contributions of this paper are summarized as follows:
\begin{itemize}
    \item We propose a timescale-separated framework decomposing the coupled optimization into instantaneous beamforming and continuous trajectory planning, addressing the inherent mismatch between channel dynamics and UAV mobility.

    \item We design a topology-aware GNN model for real-time beamforming, enabling scalable interference management across variable network densities.
    
    \item We solve the multi-UAV trajectory planning problem using MAPPO under the CTDE paradigm. To tackle sparse feedback and asynchronous termination, we design a novel reachability-aware reward mechanism with an arrival masking scheme, improving training stability and cooperative learning efficiency.

    \item Extensive simulations demonstrate the proposed framework consistently outperforms representative optimization and learning-based baselines regarding long-term system sum rate, convergence behavior, and generalization across diverse user distributions and network topologies.
\end{itemize}

The remainder of this paper is organized as follows: The related works are summarized in Section \ref{relatedwork}. 
The system model and problem formulation are introduced in Section \ref{model}. 
The GNN-based beamforming is proposed in Section \ref{gnn} and the MAPPO-based multi-UAV trajectory planning is introduced in Section \ref{mappo}. 
Simulation results are discussed in Section \ref{simulation}. 
Finally, the conclusion is summarized in Section \ref{cons}.

\section{Related Work}
\label{relatedwork}
Research on joint beamforming and trajectory optimization for UAV-assisted wireless communications has evolved from conventional optimization-driven frameworks to learning-based frameworks, and recently to graph-based frameworks.

\textbf{Optimization–Based Frameworks:} Conventional mathematical optimization extensively addresses the coupled UAV mobility and resource allocation design. 
Studies typically formulate these as non-convex problems, solving them via iterative algorithms like block coordinate descent, successive convex approximation, and heuristics across various scenarios, including millimeter-wave networks and wireless power transfer \cite{opt1,opt2,opt3,opt4}. 
Recent works on reconfigurable intelligent surfaces decompose the problem into sub-problems solved alternately via fractional programming \cite{opt5,opt6,opt7}.
Moreover, advanced methods like semidefinite relaxation and model predictive control tackle stringent constraints in emerging scenarios, including integrated sensing and communication, covert communications, and satellite-terrestrial coexistence \cite{opt10,opt11,opt13}.

Despite theoretical rigor, optimization-based methods suffer from prohibitive computational latency, poor scalability, and a reliance on static snapshots that fail to capture instantaneous topological changes in multi-UAV flight. 
These limitations motivate our scalable, real-time learning-based framework for efficient cooperative decision-making.

\textbf{Learning-Based Frameworks:} To overcome the high complexity and limited real-time adaptability of conventional optimization, RL is widely adopted for single-UAV trajectory and phase-shift designs using continuous control algorithms \cite{ml1,ml2}.
Beyond pure RL, supervised techniques like recurrent neural networks predict beam alignment in high-mobility scenarios \cite{ml3}. 
For expanding network scales, research shifted towards MARL to handle collaborative tasks, including phase synchronization and mmWave beam tracking \cite{ml4,ml5}. 
Recent advancements further integrate multi-objective algorithms, federated learning, and large language models to respectively balance conflicting metrics, preserve privacy, and improve sample efficiency \cite{ml6,ml7,ml8}.

Although efficient, standard learning methods treat the wireless environment as a simple Euclidean space, ignoring inherent graph-structured interference topologies. This structural agnosticism degrades generalization across varying network scales, necessitating a shift towards graph-based paradigms for scalable topology-aware optimization.

\textbf{Graph-Based Frameworks:} Motivated by wireless networks' graph-structured nature, GNNs explicitly capture node connectivity and interference coupling. 
Early works applied GNNs primarily for snapshot-based resource allocation, optimizing link scheduling, hypergraph-based beamforming, and 3D placement by extracting spatial and interference topologies \cite{gnn1,gnn2,gnn3}. 
Recently, hybrid frameworks combining GNNs with DRL emerged for sequential decision-making tasks like age of information management via QMIX \cite{gnn4}. 
Extending this to continuous mobility control, studies integrate graph representation learning with UAV trajectory planning, utilizing GNNs for beamforming prediction and RL for trajectory optimization \cite{gnn5,gnn6}.

Despite this progress, existing studies often assume simplified scenarios with fixed topologies and synchronous behaviors \cite{gnn5,gnn6}, ignoring dynamic user associations and evolving interference in practical multi-UAV systems. 
Furthermore, jointly optimizing beamforming and trajectory on a single timescale neglects the disparity between fast channel dynamics and slower UAV mobility, enlarging the action space and destabilizing asynchronous learning. 
To overcome these limitations, we propose a topology-adaptive and timescale-aware framework leveraging dynamic heterogeneous graphs for instantaneous beamforming and stability-enhanced multi-agent policy optimization for long-horizon trajectory planning.
\section{System Model and Problem Formulation}
\label{model}
\begin{figure}[!t]
    \centering
    \includegraphics[width=0.9\linewidth]{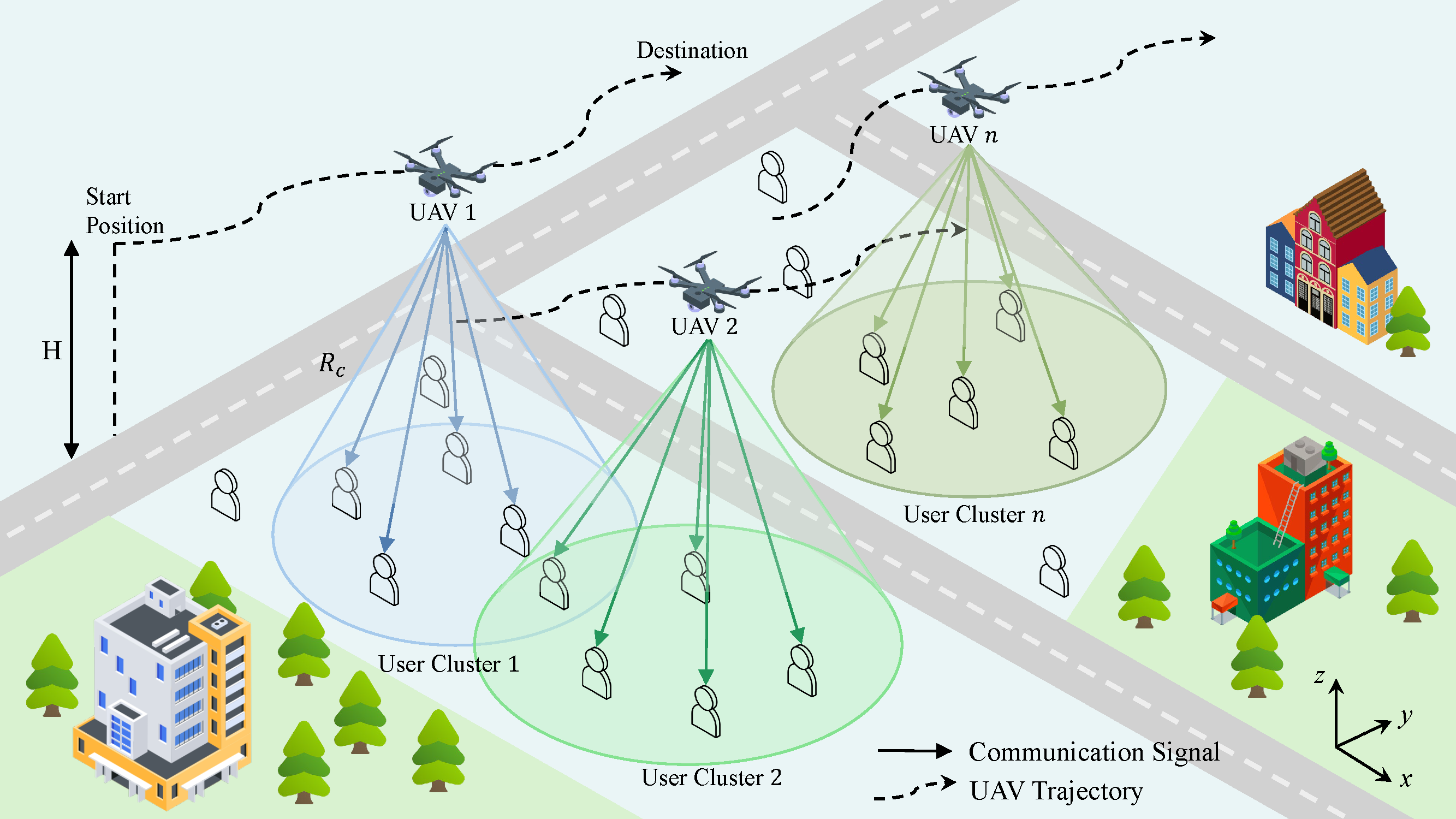}
    \caption{System model}
    \label{fig:system_model}
\end{figure}
As illustrated in Fig. \ref{fig:system_model}, we investigate a multi-UAV multi-user downlink communication network, where a set of $N$ rotary-wing UAVs, denoted by $\mathcal{N}=\{1, \ldots, N\}$, are dispatched to serve $K$ randomly distributed ground users, denoted by $\mathcal{K}=\{1, \ldots, K\}$.
Each UAV $n \in \mathcal{N}$ is equipped with a uniform linear array (ULA) of $L$ antennas, whereas each user is equipped with a single omnidirectional antenna.
We assume that all UAVs fly at a fixed altitude $H$ with a constant speed $V$.
The system operation timeline is discretized into equal-length time slots indexed by $t \in \mathcal{T} = \{1, \ldots, T\}$, where the duration of each slot is $\delta_t$.
Without loss of generality, we adopt a 3D Cartesian coordinate system.
The location of user $k$ is fixed at $\boldsymbol{l}^{\text{U}}_k = [x^\text{u}_k, y^\text{u}_k, 0]^T$.
Meanwhile, the time-varying position of UAV $n$ at time slot $t$ is denoted by $\boldsymbol{l}^{\text{A}}_n[t] = [x^\text{a}_n[t], y^\text{a}_n[t], H]^T$.
For the trajectory planning task, each UAV $n$ is required to travel from a predefined starting point $\boldsymbol{l}_n^{\mathrm{S}} = [x^{\mathrm{s}}_n, y^{\mathrm{s}}_n, H]^T$ to a destination point $\boldsymbol{l}_n^{\mathrm{D}} = [x^{\mathrm{d}}_n, y^{\mathrm{d}}_n, H]^T$ within the maximum mission duration $T_{\text{max}}$, while simultaneously providing downlink data transmission to the served users.

\subsection{Dynamic User Association and UAV Mobility}
To guarantee reliable link quality, we assume that each UAV has a limited communication coverage radius, denoted by $R_c$. 
A user $k$ is considered serviceable by UAV $n$ only if it is within the coverage region, satisfying the geometric constraint $\|\boldsymbol{l}^{\text{A}}_n[t] - \boldsymbol{l}^{\text{U}}_k\| \leq R_c$. 
To handle the overlapping coverage areas, we adopt a nearest-neighbor association protocol. 
Specifically, at any time slot $t$, a user is associated with the spatially closest UAV that satisfy the coverage constraint. 
Consequently, the user set served by UAV $n$ at time slot $t$, denoted by $\mathcal{K}_n[t]$, is referred to as a user cluster:
\begin{equation} \label{eq:cluster_association}
\begin{split}
    \mathcal{K}_n[t] = \bigg\{ k \in \mathcal{K}[t] \mid & \|\boldsymbol{l}^{\text{A}}_n[t] - \boldsymbol{l}^{\text{U}}_k\| \leq R_c, \\
    & n = \arg \min_{j \in \mathcal{N}} \|\boldsymbol{l}^{\text{A}}_j[t] - \boldsymbol{l}^{\text{U}}_k\| \bigg\}.
\end{split}
\end{equation}
Furthermore, to eliminate severe inter-cluster interference among neighboring UAVs, we assume that the total system bandwidth is orthogonally allocated to different UAVs (e.g., via frequency division multiple access).
Under this assumption, the multi-UAV network is decoupled into $N$ locally independent downlink subsystems \cite{Coverage}. 
Thus, the signal quality of each user is solely determined by the channel gain from its serving UAV and the intra-cluster interference caused by other users within the same cluster $\mathcal{K}_n[t]$.

The mobility of UAVs is modeled as a discrete-time decision-making process. 
At each time slot $t$, UAV $n$ executes a movement action $\boldsymbol{a}_n[t]$, which dictates its displacement for the subsequent time slot. 
To ensure the physical feasibility and safety of the flight trajectories, the following constraints are imposed:

\begin{itemize}
    \item \textbf{Collision Avoidance:} To prevent aerial accidents, a minimum safety distance $D_{\min}$ must be strictly maintained between any pair of UAVs throughout the mission duration:
    \begin{equation} \label{eq:collision_constraint}
        \|\boldsymbol{l}^{\text{A}}_n[t] - \boldsymbol{l}^{\text{A}}_m[t]\|^2 \geq D_{\min}^2, \quad\forall n \neq m, \forall t \in \mathcal{T}.
    \end{equation}

    \item \textbf{Flight Range Limits:} To ensure the UAVs operate within the service area, the horizontal position of each UAV $n$ is strictly bounded by the predefined rectangular region $[x_{\min}, x_{\max}] \times [y_{\min}, y_{\max}]$:
    \begin{align} 
        x_{\min} \leq x^\text{a}_n[t] \leq x_{\max}, \quad\forall n \in \mathcal{N}, \forall t \in \mathcal{T},\label{eq:range_x} \\
        y_{\min} \leq y^\text{a}_n[t] \leq y_{\max}, \quad\forall n \in \mathcal{N}, \forall t \in \mathcal{T}.\label{eq:range_y}
    \end{align}
    
    \item \textbf{Mission Duration Constraint:} Due to on-board battery limitations, each UAV $n$ is required to arrive at its destination $\boldsymbol{l}_n^{\mathrm{D}}$ within a maximum allowable flight duration $T_{\max}$. 
    Let $T$ denote the actual system operation time, defined as the maximum completion time among all UAVs. This imposes the constraint:
    \begin{equation} \label{eq:flight_time_constraint}
        T \triangleq \max_{n \in \mathcal{N}} T_n \leq T_{\max}.
    \end{equation}
\end{itemize}

\subsection{Channel Model and Signal Transmission}
Given the dynamic mobility of UAVs, the air-to-ground (A2G) channels exhibit strong time-varying characteristics. 
We adopt a Line-of-Sight (LoS) dominated channel model, which is widely used for UAV communications in high-altitude scenarios \cite{opt1}\cite{opt2}\cite{opt6}. 
At time slot $t$, the channel vector $\mathbf{h}_{n,k}[t] \in \mathbb{C}^{L \times 1}$ between UAV $n$ and user $k$ is modeled as:
\begin{equation} \label{eq:channel_model}
    \mathbf{h}_{n,k}[t] = \sqrt{\beta_{n,k}[t]} \mathbf{a}(\theta_{n,k}[t]),
\end{equation}
where $\beta_{n,k}[t] = \beta_0 (d_0 / d_{n,k}[t])^{-2}$ represents the large-scale path loss. 
Here, $d_{n,k}[t] = \|\boldsymbol{l}^{\text{A}}_n[t] - \boldsymbol{l}^{\text{U}}_k\|$ denotes the instantaneous distance, $\beta_0$ is the channel power gain at the reference distance $d_0$.
Furthermore, $\mathbf{a}(\theta_{n,k}[t])$ denotes the array steering vector. 
Assuming each UAV is equipped with a ULA with antenna spacing $d_a$, the steering vector is given by:
\begin{equation} \label{eq:steering_vector}
    \mathbf{a}(\theta_{n,k}[t]) = \left[1, e^{j 2\pi \frac{d_a}{\lambda} \cos \theta_{n,k}[t]}, \ldots, e^{j 2\pi \frac{d_a}{\lambda} (L-1) \cos \theta_{n,k}[t]} \right]^T,
\end{equation}
where $\lambda$ is the carrier wavelength, and $\theta_{n,k}[t]$ represents the elevation angle of departure (AoD), calculated as $\theta_{n,k}[t] = \arccos(H / d_{n,k}[t])$.

To support downlink transmission, UAV $n$ employs a linear precoding vector $\mathbf{w}_{n,k}[t] \in \mathbb{C}^{L \times 1}$ for each associated user $k \in \mathcal{K}_n[t]$, which is controlled by the beamforming algorithm. 
$\mathbf{W}_{n}[t]$ denotes all the beamforming vectors within the cluster $\mathcal{K}_n[t]$.
The transmitted signal $\mathbf{x}_n[t]$ is the superposition of data symbols intended for its serving cluster:
\begin{equation}
    \mathbf{x}_n[t] = \sum_{k \in \mathcal{K}_n[t]} \mathbf{w}_{n,k}[t] s_{n,k}[t],
\end{equation}
where $s_{n,k}[t] \sim \mathcal{CN}(0,1)$ is the normalized data symbol. 
The transmit power of each UAV is constrained by a maximum budget $P_{\max}$:
\begin{equation} \label{eq:power_constraint}
    \sum_{k \in \mathcal{K}_n[t]} \|\mathbf{w}_{n,k}[t]\|^2 \leq P_{\max}, \quad \forall n \in \mathcal{N}.
\end{equation}
The received signal at user $k$ (served by UAV $n$) comprises the desired signal, the intra-cluster interference from other users served by the same UAV, and the additive noise:
\begin{equation} \label{eq:received_signal}
\begin{split}
    y_k[t] = & \underbrace{\mathbf{h}_{n,k}^H[t] \mathbf{w}_{n,k}[t] s_{n,k}[t]}_{\text{Desired Signal}} \\
    & + \underbrace{\sum_{j \in \mathcal{K}_n[t], j \neq k} \mathbf{h}_{n,k}^H[t] \mathbf{w}_{n,j}[t] s_{n,j}[t]}_{\text{Intra-cluster Interference}} + z_k[t],
\end{split}
\end{equation}
where $z_k[t] \sim \mathcal{CN}(0, \sigma_0^2)$ denotes the additive white Gaussian noise (AWGN).
The achievable signal-to-interference-plus-noise ratio (SINR) is given by:
\begin{equation} \label{eq:sinr}
    \gamma_{k}[t] = \frac{|\mathbf{h}_{n,k}^H[t] \mathbf{w}_{n,k}[t]|^2}{\sum_{j \in \mathcal{K}_n[t], j \neq k} |\mathbf{h}_{n,k}^H[t] \mathbf{w}_{n,j}[t]|^2 + \sigma_0^2}.
\end{equation}
Accordingly, the achievable data rate of user $k$ at time slot $t$ is given by:
\begin{equation} \label{eq:rate}
    R_{n,k}[t] = \log_2(1 + \gamma_{n,k}[t]).
\end{equation}

\subsection{Problem Formulation}
Our primary objective is to maximize the long-term average sum rate $\bar{R}$ by jointly optimizing the active beamforming vectors $\mathcal{W} \triangleq \{\mathbf{w}_{n,k}[t]\}$ and the UAV trajectories $\mathcal{Q} \triangleq \{\boldsymbol{l}^{\text{A}}_n[t]\}$ over the entire mission duration. 
The joint optimization problem is mathematically formulated as follows:

\begin{subequations} \label{eq:P1}
\begin{align}
    (\mathbf{P1}): \quad & \max_{\mathcal{Q}, \mathcal{W}} \quad \bar{R}=\frac{1}{T} \sum_{t\in \mathcal{T}} \sum_{n \in \mathcal{N}} \sum_{k \in \mathcal{K}_n[t]} R_{n,k}(\boldsymbol{l}^{\text{A}}_{n}[t], \mathbf{W}_{n}[t]) \label{eq:P1_obj} \\
    \text{s.t.} \quad & \sum_{k \in \mathcal{K}_n[t]} \|\mathbf{w}_{n,k}[t]\|^2 \leq P_{\max}, \quad \forall n \in \mathcal{N}, \forall t \in \mathcal{T}, \label{eq:P1_power} \\
    & \boldsymbol{l}^{\text{A}}_{n}[1] = \boldsymbol{l}_{n}^{\mathrm{S}}, \quad \boldsymbol{l}^{\text{A}}_{n}[T] = \boldsymbol{l}_{n}^{\mathrm{D}}, \quad \forall n \in \mathcal{N}, \label{eq:P1_boundary} \\
    & \text{Constraints } \eqref{eq:collision_constraint}, \eqref{eq:range_x}, \eqref{eq:range_y}, \text{and } \eqref{eq:flight_time_constraint}. \label{eq:P1_time}
\end{align}
\end{subequations}
Constraint \eqref{eq:P1_power} limits the instantaneous transmit power of each UAV,
and constraint \eqref{eq:P1_boundary} enforce the start/end locations of the UAVs.

\subsection{Problem Decomposition}
\begin{figure*}[t]
    \centering
    \includegraphics[width=0.95\textwidth]{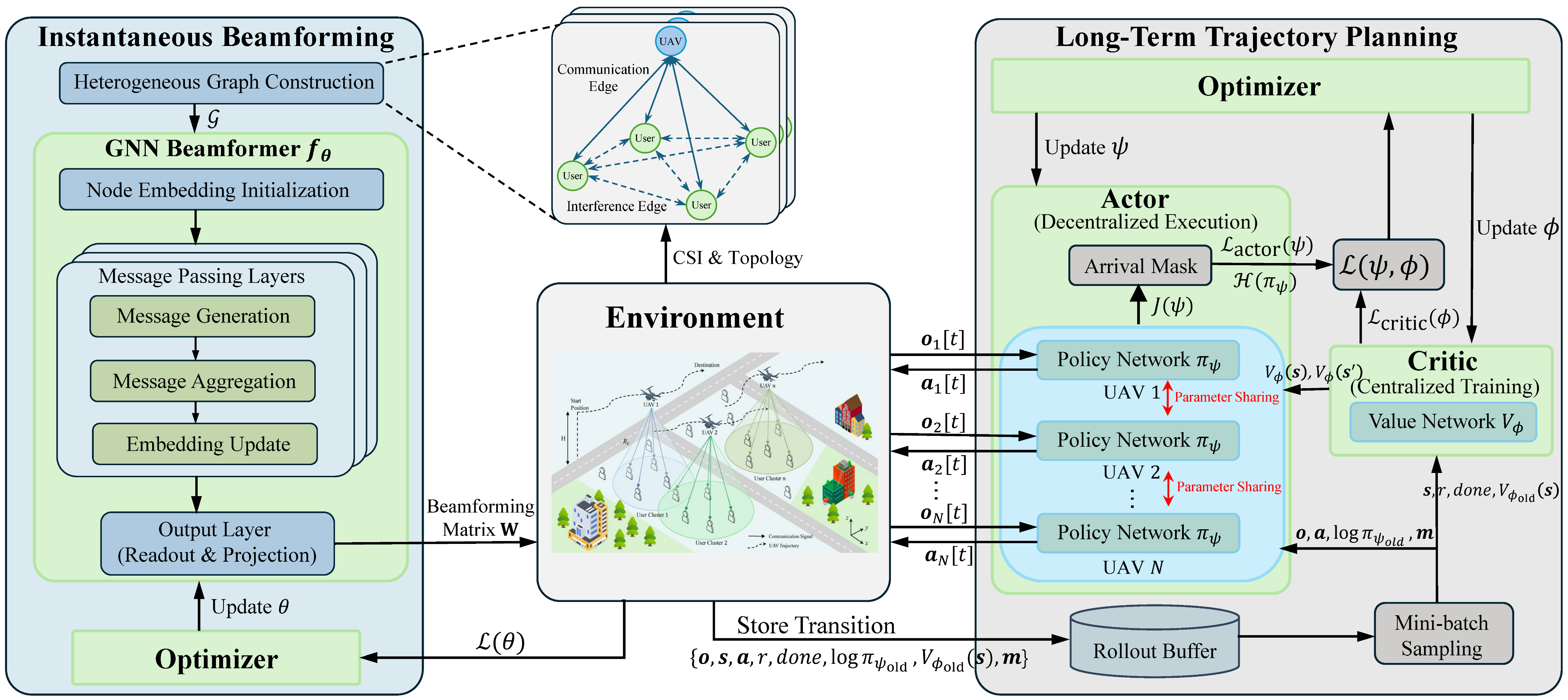}
    \caption{Overall framework of the proposed GNN-enabled beamforming and MAPPO-based UAV trajectory optimization.}
    \label{framework}
\end{figure*}

Problem $(\mathbf{P1})$ poses significant challenges due to its mixed-integer non-convex nature. 
The complex coupling between trajectory variables $\boldsymbol{l}^{\text{A}}_n[t]$ and beamforming vectors $\mathbf{w}_{n,k}[t]$, combined with the position-dependent user association $\mathcal{K}_n[t]$ determined via a nearest-neighbor rule, introduces discontinuities with respect to $\boldsymbol{l}^{\text{A}}_n[t]$.
This makes the objective function non-differentiable and difficult to optimize using standard gradient methods.
To address this, we exploit the inherent timescale separation between millisecond-level channel variations and second-level UAV mobility.
Consequently, we decompose $(\mathbf{P1})$ into two hierarchical subproblems: an inner-loop instantaneous beamforming optimization $(\mathbf{P2.1})$ and an outer-loop long-term trajectory planning $(\mathbf{P2.2})$. 
The overall proposed solution framework is illustrated in Fig. \ref{framework}.

For the fixed UAV locations $\boldsymbol{l}^{\text{A}}_n[t]$ at time slot $t$, the inner-loop subproblem $(\mathbf{P2.1})$ maximizes the instantaneous sum rate $R[t]$ by optimizing the beamforming vectors $\mathcal{W}$:
\begin{subequations} \label{eq:P2}
\begin{align}
    (\mathbf{P2.1}): \quad & R[t]=\max_{\mathcal{W}} \quad \sum_{n \in \mathcal{N}} \sum_{k \in \mathcal{K}_n[t]} R_{n,k}(\boldsymbol{l}^{\text{A}}_n[t],\mathbf{W}_{n}[t]) \label{eq:P2_obj} \\
    \text{s.t.} \quad & \sum_{k \in \mathcal{K}_n[t]} \|\mathbf{w}_{n,k}[t]\|^2 \leq P_{\max}, \quad \forall n \in \mathcal{N}. \label{eq:P2_power}
\end{align}
\end{subequations}
Since $(\mathbf{P2.1})$ remains non-convex and dominated by intra-cluster interference, conventional iterative solvers incur prohibitive latency for real-time channels. Thus, we propose a topology-aware GNN beamformer to approximate the optimal solution with low inference latency.

The outer-loop subproblem $(\mathbf{P2.2})$ optimizes the UAV trajectories $\mathcal{Q}$ over the mission duration to maximize the long-term average sum rate $\overline{R}$. Given the beamforming vectors generated by the pre-trained GNN, 
i.e.,
$\mathbf{W}_n[t] = f_{\theta}(\mathbf{H}[t])$. 
$(\mathbf{P2.2})$ is formulated as:
\begin{subequations} \label{eq:P3}
\begin{align}
    (\mathbf{P2.2}): \quad & \bar{R}=\max_{\mathcal{Q}} \quad \frac{1}{T} \sum_{t \in \mathcal{T}} \sum_{n \in \mathcal{N}} \sum_{k \in \mathcal{K}_n[t]} R_{n,k}\big(\boldsymbol{l}^{\text{A}}_n[t], f_{\theta}(\mathbf{H}[t])) \label{eq:P3_obj} \\
    \text{s.t.} \quad & \boldsymbol{l}^{\text{A}}_n[1] = \boldsymbol{l}_n^{\mathrm{S}}, \quad \boldsymbol{l}^{\text{A}}_n[T] = \boldsymbol{l}_n^{\mathrm{D}}, \quad \forall n \in \mathcal{N}, \label{eq:P3_boundary} \\
    &  \text{Constraints } \eqref{eq:collision_constraint}, \eqref{eq:range_x}, \eqref{eq:range_y}, \text{and } \eqref{eq:flight_time_constraint}. \label{eq:P1_time}
\end{align}
\end{subequations}
To tackle the exponential complexity of this constrained sequential decision-making process, we reformulate $(\mathbf{P2.2})$ as a MARL task solved via MAPPO.

As shown in Fig. \ref{framework}, the proposed framework integrates the two modules into a closed loop. 
The MAPPO acts as the high-level planner, determining the UAV movements for the next time step, while the GNN beamformer acts as part of the environment, rapidly calculating the optimal beamforming strategy and the resulting achievable sum rate. 
This rate provides immediate reward signals to guide MAPPO training, enabling joint communication and control optimization.

\section{GNN-BASED BEAMFORMING}
\label{gnn}
This section details the inner-loop solution for the instantaneous beamforming subproblem $(\mathbf{P2.1})$. 
\subsection{Heterogeneous Graph Construction}
At each time slot $t$, the local communication environment of UAV $n$ is modeled as a time-varying heterogeneous graph $\mathcal{G}_n[t] = (\mathcal{V}_n[t], \mathcal{E}_n[t])$. 
The node set $\mathcal{V}_n[t]$ comprises a central UAV node $v_n$ and $K_n[t] = |\mathcal{K}_n[t]|$ associated user nodes $\{u_k\}_{k \in \mathcal{K}_n[t]}$. 
The edge set $\mathcal{E}_n[t]$ forms a fully connected topology within the cluster, where UAV-user edges represent desired communication links and user-user edges $(u_k, u_j)$ capture potential intra-cluster interference coupling. 
This unified structure enables the GNN to explicitly encode the interference structure induced by spatial proximity and shared transmission resources. 
As optimization occurs per time slot, each constitutes an independent graph instance. 
Crucially, a single parameter-shared GNN processes all instances, allowing the framework to generalize across varying network topologies and cluster sizes $K_n[t]$.

To ensure transferability across geometric configurations, we design node features using solely instantaneous local CSI. 
Specifically, decomposing the complex downlink channel vector $\mathbf{h}_{n,k}[t] \in \mathbb{C}^{L \times 1}$ into real and imaginary parts yields the initial feature vector $\mathbf{x}_k^{(0)}[t] \in \mathbb{R}^{2L}$ for user node $u_k$ \cite{gnn6}:
\begin{equation} \label{eq:node_feature}
    \mathbf{x}_k^{(0)}[t] = \left[ \Re\{\mathbf{h}_{n,k}[t]\}^T, \; \Im\{\mathbf{h}_{n,k}[t]\}^T \right]^T.
\end{equation}
By excluding explicit geometric data (e.g., AoD or user coordinates), this CSI-based representation remains independent of specific deployment layouts. 
Combined with the GNN's permutation-invariant aggregation, this design improves robustness and generalization across varying user distributions and cluster sizes.

\subsection{GNN Architecture with GraphNorm}
To effectively handle the time-varying heterogeneous graph size caused by the UAV mobility, we design a GraphNorm-enhanced GNN architecture \cite{GraphNorm}. 
Unlike Batch Normalization, which relies on batch statistics, GraphNorm normalizes node features based on the statistics of the current graph instance \cite{GraphNorm}, thereby improving robustness across varying cluster sizes.
We integrate GraphNorm into all multi-layer perceptron (MLP) blocks within the GNN model. 

Let $\mathbf{H} = [\mathbf{h}_1, \dots, \mathbf{h}_K]^T \in \mathbb{R}^{K \times D}$ denote the matrix of node features in a graph instance, where $K$ is the number of nodes and $D$ is the feature dimension. 
The GNN normalized along the node dimension:
\begin{equation} \label{eq:graph_norm_def}
    \text{GN}(\mathbf{h}_{n,k}) = \boldsymbol{\gamma} \odot \frac{\mathbf{h}_{n,k} - \boldsymbol{\mu}}{\sqrt{\boldsymbol{\sigma}^2 + \epsilon}} + \boldsymbol{\beta},
\end{equation}
where $\boldsymbol{\mu}$ and $\boldsymbol{\sigma}^2$ are the mean and variance computed over the $K$ nodes in the current graph, and $\boldsymbol{\gamma}, \boldsymbol{\beta}$ are learnable affine parameters. 
This normalization stabilizes the feature distribution regardless of the instantaneous user number $K_n[t]$ across graphs with different sizes.

As shown in Fig. \ref{framework}, the architecture of the GNN model follows the standard message passing paradigm. The detailed procedures are described as follows:

\textbf{1) Node Embedding Initialization:} The raw CSI features $\mathbf{x}_k^{(0)}$ are projected into high-dimensional user node embeddings $\mathbf{u}_k^{(0)}$  using a shared user encoder $\text{MLP}^{(u)}_{\text{enc}}$. 
The UAV node embedding $\mathbf{a}_n^{(0)}$ is then initialized by aggregating the normalized user embeddings using UAV encoder $\text{MLP}^{(a)}_{\text{enc}}$:
\begin{equation}
    \mathbf{u}_k^{(0)} = \text{MLP}^{(u)}_{\text{enc}}(\mathbf{x}_k^{(0)})\in\mathbb{R}^{d_u},
\end{equation}
\begin{equation}
    \mathbf{a}_n^{(0)} = \text{MLP}^{(a)}_{\text{enc}}\left( \frac{\max(1,K)}{K_n[t]} \sum_{k \in \mathcal{K}_n[t]} \mathbf{u}_k^{(0)} \right)\in\mathbb{R}^{d_a},
\end{equation}
where $d_u = d_a = d_{\text{gnn}}$ denote the embedding dimensions of the user nodes and UAV nodes.

\textbf{2) Message Passing Process:} The GNN model consists of $L_g$ message-passing layers. In the $\ell$-th layer, messages of user and UAV nodes are generated using their corresponding message-generation modules:
\begin{equation}
    \mathbf{m}_{\mathbf{u}_k}^{(\ell)} = \text{MLP}_{\text{msg}}^{(u)}(\mathbf{u}_k^{(\ell-1)})\in\mathbb{R}^{d_u}, 
\end{equation}
\begin{equation}
    \mathbf{m}_{\mathbf{a}_n}^{(\ell)} = \text{MLP}_{\text{msg}}^{(a)}(\mathbf{a}_n^{(\ell-1)})\in\mathbb{R}^{d_a}.
\end{equation}
User nodes aggregate interference information via max pooling to capture the dominant interferer and combine it with the UAV's global guidance:
\begin{equation}
    \bar{\mathbf{m}}_{\mathbf{u}_k}^{(\ell)} = \text{concat}\left( \mathbf{m}_{\mathbf{a}_n}^{(\ell)}, \max_{j \in \mathcal{K}_n[t], j \neq k} \mathbf{m}_{\mathbf{u}_j}^{(\ell)} \right)\in \mathbb{R}^{d_u+d_a}.
\end{equation}
For the UAV node, mean pooling is employed to aggregate the messages from all associated user nodes:
\begin{equation}
    \bar{\mathbf{m}}_{\mathbf{a}_n}^{(\ell)}=\frac{1}{K_n[t]}\sum_{k\in\mathcal{K}_n[t]}\mathbf{m}_{\mathbf{u}_k}^{(\ell)}\in\mathbb{R}^{d_u}.
\end{equation}
Finally, the embeddings of both the user and UAV nodes are updated using the updater $\text{MLP}_{\text{upd}}^{(u)}$ and $\text{MLP}_{\text{upd}}^{(a)}$ based on the aggregated messages:
\begin{equation}
    \mathbf{u}_k^{(\ell)}=\text{MLP}_{\text{upd}}^{(u)}\left(\mathbf{u}_k^{(\ell-1)},\bar{\mathbf{m}}_{\mathbf{u}_k}^{(\ell)}\right)\in\mathbb{R}^{d_u},
\end{equation}
\begin{equation}
    \quad\mathbf{a}_n^{(\ell)}=\text{MLP}_{\text{upd}}^{(a)}\left(\mathbf{a}_n^{(\ell-1)},\bar{\mathbf{m}}_{\mathbf{a}_n}^{(\ell)}\right)\in\mathbb{R}^{d_a}.
\end{equation}
This message-passing enables learning of intra-cluster interference patterns without explicitly constructing edge features.

\textbf{3) Beamforming Vector Output:} The final user embedding $\mathbf{u}_k^{(L_g)}$ is passed through a readout head to generate the beamforming vector $\tilde{\mathbf{w}}_{n,k}$:
\begin{equation}
    \tilde{\mathbf{w}}_{n,k} = \text{MLP}_{\text{out}}(\mathbf{u}_k^{(L_g)}).
\end{equation}
To enforce the per-UAV transmit power constraint, the beamforming vectors are scaled as:
\begin{equation}
    \alpha_n(t)=\min\left(1,\sqrt{\frac{P_{\max}}{\sum_{k\in\mathcal{K}_n(t)}\|\tilde{\mathbf{w}}_{n,k}(t)\|^2+\varepsilon}}\right),
\end{equation}
\begin{equation}
    \mathbf{w}_{n,k}(t)=\alpha_n(t)\tilde{\mathbf{w}}_{n,k}(t),
\end{equation}
where $\varepsilon>0$ is a numerical stability term introduced to avoid numerical instability or gradient explosion when the total unnormalized beamforming power approaches zero. This projection operation guarantees that the output beamforming vectors always satisfy the transmit power constraint while preserving end-to-end differentiability.

\subsection{Training and Complexity Analysis}
\label{gnn_complexity}
Instead of relying on labeled data generated by computationally expensive iterative solvers, we adopt a reward-driven learning paradigm that aligns with the principles of RL.
In this context, the GNN beamformer $f_{\boldsymbol{\theta}}$ functions as a policy network, which directly maps the environmental state (i.e., the CSI matrix) to the optimal action (i.e., beamforming vectors). 
The network parameters $\boldsymbol{\theta}$ are optimized via stochastic gradient descent (SGD) to maximize the instantaneous reward, defined as the system sum rate.
Formally, let $\mathcal{D}_{\text{gnn}}$ denote the training dataset consisting of clustered channel samples. During the training phase, we sample a mini-batch $\mathcal{B}_{\text{gnn}}$ of size $B_{\text{gnn}}$. Since each user cluster constitutes an independent graph instance, the training objective is to maximize the average expected reward $\bar{R}_{\text{clus}}(\boldsymbol{\theta})$ over the mini-batch, which is defined as:
\begin{equation} \label{eq:cluster sum rate}
    \bar{R}_{\text{clus}}(\boldsymbol{\theta}) = \frac{1}{B_{\text{gnn}}\times N} \sum_{m \in \mathcal{B}_{\text{gnn}}} \sum_{n\in\mathcal{N}} \sum_{k \in \mathcal{K}_{m,n}} R_k(\mathbf{W}_{m,n}),
\end{equation}
where $\mathbf{W}_{m,n}$ is the beamforming matrix generated by $\mathbf{W}_{m,n} = f_{\boldsymbol{\theta}}(\mathbf{H}_{m,n})$ and $\mathbf{H}_{m,n}$ represents the CSI matrix of the $n$-th cluster in sample $m$. The loss function is defined as:
\begin{equation} \label{eq:loss_function}
    \mathcal{L}(\boldsymbol{\theta}) = - \bar{R}_{\text{clus}}(\boldsymbol{\theta}),
\end{equation}

The detailed training procedure is outlined in Algorithm \ref{alg:gnn-training}. 
To enhance training efficiency, a cluster-level batching strategy is employed. 
Specifically, since the interference is localized within each cluster, multiple clusters from different time slots are flattened into a single batch. 
This allows the GNN to process diverse topology instances in parallel, significantly improving sample efficiency and convergence speed. 

The computational complexity is dominated by the message passing process. 
Node updates incur a cost of $\mathcal{O}(L_g (N+K) d_{\text{gnn}}^2)$, while message aggregation scales with the total number of edges $|\mathcal{E}|$ as $\mathcal{O}(L_g |\mathcal{E}| d_{\text{gnn}})$. 
Since the edges are strictly restricted to intra-cluster connections with a limited cluster size $K_n$, $|\mathcal{E}|$ scales linearly with the total user number $K$ (i.e., $|\mathcal{E}| \propto K$). 
This has an overall complexity of $\mathcal{O}(K)$, ensuring scalability.

\begin{algorithm}[t]
\caption{GNN Training}
\label{alg:gnn-training}
\renewcommand{\algorithmicrequire}{\textbf{Input:}}
\renewcommand{\algorithmicensure}{\textbf{Output:}}
\begin{algorithmic}[1]
\REQUIRE Clustered dataset $\mathcal{D}_{\text{gnn}}$, UAV number $N$, batch size $B_{\text{gnn}}$, epoch number $E_{\text{gnn}}$, 
         GNN beamformer $f_{\boldsymbol{\theta}}$;
\ENSURE Trained parameters $\boldsymbol{\theta}$;

\STATE Initialize GNN beamformer $f_{\boldsymbol{\theta}}$;
\STATE Load clustered dataset $\mathcal{D}_{\text{gnn}}=\{(\mathcal{C}_m)\}_{m=1}^{M}$ where each sample $m$ has clusters $\mathcal{C}_m = \{\mathbf{H}_{m,n}\}_{n=1}^N$;

\FOR{epoch $=1$ to $E_{\text{gnn}}$}
    \STATE Shuffle sample indices;
    \FOR{each mini-batch $\mathcal{B}_{\text{gnn}}$}
        \STATE Extract all clusters in $\mathcal{B}_{\text{gnn}}$ into list $\mathcal{C}$; 
        \STATE $L_{\text{sum}} \leftarrow 0$, $C_{\text{cnt}} \leftarrow 0$;
        \FOR{each cluster in $\mathcal{C}$}
            \STATE Output the beamforming matrix with $f_{\boldsymbol{\theta}}$: 
            \STATE $\mathbf{W}_{m,n} \leftarrow f_{\boldsymbol{\theta}}(\mathbf{H}_{m,n})$; 
            \STATE Compute cluster sum-rate loss $\mathcal{L}_{m,n}$;
            \STATE Update $\mathcal{L}_{\text{sum}}, C_{\text{cnt}}$:
            \STATE $\mathcal{L}_{\text{sum}} \leftarrow \mathcal{L}_{\text{sum}} + \mathcal{L}_{m,n}$, $C_{\text{cnt}} \leftarrow C_{\text{cnt}} + 1$;
        \ENDFOR
        \STATE $\mathcal{L}(\boldsymbol{\theta}) \leftarrow \mathcal{L}_{\text{sum}} / C_{\text{cnt}}$;
        \STATE Update $\boldsymbol{\theta}$ with $\mathcal{L}_{\text{batch}}$:
        \STATE $\boldsymbol{\theta}\leftarrow\boldsymbol{\theta}-\eta\nabla_{\boldsymbol{\theta}} L(\boldsymbol{\theta})$;
    \ENDFOR
\ENDFOR
\end{algorithmic}
\end{algorithm}

\section{MAPPO-BASED MULTI-UAV TRAJECTORY PLANNING}
This section elaborates on the outer-loop solution for the long-term trajectory planning subproblem $(\mathbf{P2.2})$. 

\label{mappo}
\subsection{Utilization of Beamforming Results}
As shown in Fig. \ref{framework}, the pre-trained GNN beamformer is integrated into the environment to provide real-time performance feedback. 
Specifically, at each time step $t$, for any given UAV locations, the GNN maps the current CSI matrix $\mathbf{H}[t]$ to the beamforming vectors $\mathbf{W}[t]$ by $\mathbf{W}[t] = f_{\boldsymbol{\theta}}(\mathbf{H}[t])$.
Subsequently, the system calculates the instantaneous sum rate $R[t]$ based on $\mathbf{W}[t]$, which serves as part of the reward signal for the MARL agents (detailed in Section \ref{Dec-POMDP}).
This mechanism encapsulates the underlying interference management, allowing the MAPPO agents to optimize long-term trajectories based solely on the observed states and the GNN-feedback rewards.

\subsection{Dec-POMDP Modeling}
\label{Dec-POMDP}
We formulate the cooperative multi-UAV trajectory planning problem as a Dec-POMDP, defined by the tuple $\mathcal{M} = \langle \mathcal{N}, \mathcal{S}, \{\mathcal{A}_n\}, \{\mathcal{O}_n\}, \mathcal{P}, \mathcal{R}, \gamma \rangle$.
Here, $\mathcal{N}$ is the set of UAV agents. At time slot $t$, $\boldsymbol{s}[t] \in \mathcal{S}$ denotes the global state. 
Each agent $n$ selects an action $\boldsymbol{a}_n[t] \in \mathcal{A}_n$ based on its local observation $\boldsymbol{o}_n[t] \in \mathcal{O}_n$. 
$\mathcal{P}$ is the state transition probability, $\mathcal{R}$ is the reward function, and $\gamma$ is the discount factor. 
The detailed components are defined as follows:

\subsubsection{Global State}
The global state $\boldsymbol{s}[t]$ encapsulates the geometric configuration of the entire network, including the locations of all UAVs, users, and destinations. It is defined as:
\begin{equation}
    \boldsymbol{s}[t] = \left\{ \{\boldsymbol{l}^{\text{A}}_n[t]\}_{n \in \mathcal{N}}, \{\boldsymbol{l}^{\text{U}}_k\}_{k \in \mathcal{K}}, \{\boldsymbol{l}^D_n\}_{n \in \mathcal{N}} \right\}.
\end{equation}
This global information is exclusively available to the Critic network during the centralized training phase to facilitate value function estimation.

\subsubsection{Local Observation}
To enable decentralized execution, the local observation $\boldsymbol{o}_n[t]$ of UAV $n$ relies solely on local information. 
It consists of its own position, the relative locations of users and the destination, and historical action information:
\begin{equation}
    \boldsymbol{o}_n[t] = \left\{ \boldsymbol{l}^{\text{A}}_n[t], \; \Delta \boldsymbol{l}_{n, \text{avg}}^{\text{U}}[t], \; \Delta \boldsymbol{l}_{n}^{\text{D}}[t], \; \boldsymbol{a}_{n}[t-1] \right\}.
\end{equation}
Here, $\Delta \boldsymbol{l}_{n, \text{avg}}^{\text{U}}[t]$ represents the average relative displacement to all ground users, calculated as $\frac{1}{K} \sum_{k \in \mathcal{K}} (\boldsymbol{l}^{\text{U}}_k - \boldsymbol{l}^{\text{A}}_n[t])$. 
This compact representation provides a guide toward user-dense regions without expanding the observation dimension as $K$ increases.
$\Delta \boldsymbol{l}_{n}^{\text{D}}[t] \triangleq \boldsymbol{l}^{\text{D}}_n - \boldsymbol{l}^{\text{A}}_n[t]$ is the vector pointing to the destination.
Additionally, the one-hot encoded previous action $\boldsymbol{a}_{n}[t-1]$ is included to promote trajectory smoothness.

\subsubsection{Action Space}
We adopt a discrete action space for UAV $n$, which is defined as:
\begin{equation}
    \mathcal{A}_n = \{ \text{UP}, \text{DOWN}, \text{LEFT}, \text{RIGHT}, \text{STAY} \}.
\end{equation}
These actions correspond to planar displacement vectors $\{(\pm D_{\text{fly}}, 0, 0), (0, \pm D_{\text{fly}}, 0), (0, 0, 0)\}$, where $D_{\text{fly}} \triangleq V \delta_t$ denotes the step size determined by the fly speed. 
The position update rule is given by $\boldsymbol{l}^{\text{A}}_n[t+1] = \boldsymbol{l}^{\text{A}}_n[t] + \boldsymbol{a}_n[t]$.

\subsubsection{Reward Function}
The reward function is meticulously designed to guide the agents toward a balance between maximizing system sum rate and ensuring safe, timely mission completion. 
The global reward $r[t]$ at time slot $t$ is formulated as a weighted sum of heterogeneous objectives:
\begin{equation} \label{eq:total_reward}
\begin{split}
    r[t] = & w_{\text{rate}} \bar{R}[t] + w_{\text{arr}} N_{\text{arr}}[t] - w_{\text{bud}} N_{\text{bud}}[t] \\
    & - w_{\text{col}} N_{\text{col}}[t] + w_{\text{fea}} \sum_{n \in \mathcal{N}} \Phi_n[t].
\end{split}
\end{equation}
Here, the integer terms $N_{\text{arr}}[t]$, $N_{\text{bud}}[t]$, and $N_{\text{col}}[t]$ represent the number of UAVs that have newly arrived at destinations, violated boundary constraints, or incurred collisions at the current slot, respectively. 
The coefficients $w_{(\cdot)}$ are hyperparameters governing the relative importance of these factors.

To prioritize long-term network performance, the communication component $\bar{R}[t]$ tracks the cumulative average sum rate achieved up to time $t$:
\begin{equation}
    \bar{R}[t] = \frac{1}{t} \sum_{\tau=1}^{t} R[\tau].
\end{equation}

Standard trajectory planning suffers from sparse rewards, as agents receive positive feedback only upon reaching the destination, making intermediate decisions difficult to optimize.
To address this, the proposed reachability term $\Phi_n[t]$ provides dense, step-wise supervision by immediately checking physical feasibility at every time slot, effectively guiding the agent to correct its trajectory continuously throughout the flight. 
Let $T_n^{\min}[t] = \lceil \|\boldsymbol{l}^{\text{D}}_n - \boldsymbol{l}^{\text{A}}_n[t]\| / D_{\text{fly}} \rceil$ denote the minimum time slots required to reach the destination from the current position, and let $T_n^{\text{res}}[t] = T_{\max} - t$ be the remaining time. 
The feasibility indicator is defined as:
\begin{equation}
    \Phi_n[t] = 
    \begin{cases}
    +1, & \text{if } T_n^{\min}[t] \le T_n^{\text{res}}[t] \land T_n^{\min}[t] \le T_n^{\min}[t-1], \\
    -1, & \text{otherwise}.
    \end{cases}
\end{equation}
This mechanism rewards agents for maintaining a feasible trajectory while approaching the target, and imposes penalties when deviations make the mission theoretically impossible to complete within the remaining time.

\subsection{MAPPO with CTDE Framework}
\subsubsection{Network Architecture} 
As shown in Fig. \ref{framework}, we implement a parameter-sharing Actor-Critic architecture, where all UAV agents share the same set of parameters for both the policy network (Actor) and the value network (Critic). 
This design not only significantly reduces the model complexity but also promotes the learning of cooperative behaviors by aggregating experiences from all agents.

\textbf{Decentralized Actor (Execution Phase)}: The Actor network, parameterized by $\psi$, serves as the local execution policy for each UAV. It maps the local observation to a stochastic action distribution, which is denoted as $\pi_\psi(\boldsymbol{a}_{n}[t] \mid \boldsymbol{o}_{n}[t])$.
Crucially, the policy depends solely on the local observation $\boldsymbol{o}_n[t]$. 
This ensures that during the flight phase, each UAV operates autonomously without requiring real-time global information exchange, thereby enabling practical distributed deployment with minimal communication overhead.

\textbf{Centralized Critic (Training Phase)}: The Critic network, parameterized by $\phi$, is employed exclusively during the offline training phase to estimate the state value function, which is denoted as $V_\phi(\boldsymbol{s}[t])$.
Unlike the Actor, the Critic is conditioned on the global state $\boldsymbol{s}[t]$,
which provides a stable and accurate return estimation for credit assignment, effectively mitigating the non-stationarity issue and stabilizing the gradient updates.

\subsubsection{Optimization Objectives}
The network parameters $\psi$ and $\phi$ are updated using the MAPPO algorithm.

\textbf{Generalized Advantage Estimation (GAE):}
To reduce the variance of gradient estimation while maintaining an acceptable bias, we employ GAE to compute the advantage function. 
The temporal-difference (TD) error $\delta_t$ at time step $t$ is defined as:
\begin{equation}
    \delta_t = r[t] + \gamma(1-done[t+1]) V_{\boldsymbol{\phi}}(\boldsymbol{s}[t+1]) - V_{\boldsymbol{\phi}}(\boldsymbol{s}[t]),
\end{equation}
where $done[t+1] \in \{0, 1\}$ serves as a termination mask that zeros out the discounted future value if the episode concludes at step $t+1$. The advantage estimate $\hat{A}_t$ is given by the exponentially weighted sum of TD errors:
\begin{equation}
    \hat{A}_t = \sum_{l=0}^{\infty} (\gamma \lambda)^l \delta_{t+l},
\end{equation}
where $\lambda \in [0, 1]$ is the GAE smoothing factor governing the bias-variance trade-off.

\textbf{Masked Actor Loss:}
The Actor network $\pi_{\boldsymbol{\psi}}$ is updated by maximizing the clipped surrogate objective. 
However, in the multi-UAV scenario, agents may reach their destinations at different time steps. 
Once a UAV completes its task, its subsequent actions (typically STAY) are deterministic and should not influence the policy gradient. 
To address this, we introduce a binary active agent mask $m_{n,t} \in \{0, 1\}$, where $m_{n,t}=0$ indicates that UAV $n$ has finished its mission.
First, let $J_{n,t}(\boldsymbol{\psi})$ denote the standard clipped surrogate objective for a specific agent $n$ at time step $t$:
\begin{equation}
    J_{n,t}(\boldsymbol{\psi}) = \min \left( \rho_{n,t} \hat{A}_{n,t}, \; \text{clip}(\rho_{n,t}, 1-\epsilon, 1+\epsilon) \hat{A}_{n,t} \right),
\end{equation}
where $\rho_{n,t}(\psi) = \frac{\pi_{\psi}(\boldsymbol{a}_{n}[t]|\boldsymbol{o}_{n}[t])}{\pi_{\psi_{\text{old}}}(\boldsymbol{a}_{n}[t]|\boldsymbol{o}_{n}[t])}$ is the probability ratio and $\epsilon$ is the clipping parameter.
Based on this, the final masked actor loss is defined as the negative average of these objectives over active agents in the mini-batch $\mathcal{B}_{\text{ppo}}$:
\begin{equation} \label{eq:actor_loss}
    \mathcal{L}_{\text{actor}}(\boldsymbol{\psi}) = - \frac{1}{\sum_{n \in \mathcal{N}} \sum_{t \in \mathcal{B}_{\text{ppo}}} m_{n,t}} \sum_{n \in \mathcal{N}} \sum_{t \in\mathcal{B}_{\text{ppo}}} m_{n,t} \cdot J_{n,t}(\boldsymbol{\psi}).
\end{equation}
This masking mechanism ensures that the policy optimization focuses exclusively on meaningful decision-making processes, effectively filtering out noise from completed agents.

\textbf{Critic Loss and Total Objective:}
The critic network $V_{\boldsymbol{\phi}}$ is updated to minimize the mean squared error between the value prediction and the estimated return. 
To prevent destructive updates caused by excessive changes in value estimation, the clipped value loss technique is adopted.
Let $\hat{R}_t = \hat{A}_t + V_{\boldsymbol{\phi}_{\text{old}}}(\boldsymbol{s}[t])$ denote the target return. 
The clipped value prediction $\tilde{V}_{\phi, t}$ is explicitly formulated to restrict the new value estimate within a trusted region:
\begin{equation}
    \tilde{V}_{\phi, t} = V_{\boldsymbol{\phi}_{\text{old}}}(\boldsymbol{s}[t]) + \text{clip}\left( V_{\boldsymbol{\phi}}(\boldsymbol{s}[t]) - V_{\boldsymbol{\phi}_{\text{old}}}(\boldsymbol{s}[t]), -\epsilon, \epsilon \right).
\end{equation}
The critic loss is then calculated as the maximum of the unclipped and clipped squared errors:
\begin{equation} \label{eq:critic_loss}
\begin{split}
    \mathcal{L}_{\text{critic}}(\boldsymbol{\phi}) = \frac{1}{|\mathcal{B}_{\text{ppo}}|} \sum_{t \in \mathcal{B}_{\text{ppo}}} \max \Big[ & \left(V_{\boldsymbol{\phi}}(\boldsymbol{s}[t]) - \hat{R}_t\right)^2, \\
    & \left(\tilde{V}_{\phi, t} - \hat{R}_t\right)^2 \Big].
\end{split}
\end{equation}
Finally, the total optimization objective is constructed as a weighted combination of the actor loss, critic loss, and entropy regularization:
\begin{equation} \label{eq:total_loss}
    \mathcal{L}_{\text{total}} (\psi,\phi)= \mathcal{L}_{\text{actor}}(\boldsymbol{\psi}) + c_1 \mathcal{L}_{\text{critic}}(\boldsymbol{\phi}) - c_2 \mathcal{H}(\pi_{\boldsymbol{\psi}}),
\end{equation}
where $c_1$ and $c_2$ represent the value loss coefficient and entropy coefficient, respectively. The entropy term $\mathcal{H}(\pi_{\boldsymbol{\psi}})$ is incorporated to encourage exploration by preventing the policy from becoming deterministic too early, thereby mitigating the risk of premature convergence to suboptimal local optima.

\subsection{Training and Complexity Analysis}
Summarized in Algorithm \ref{alg:mappo_training}, the training workflow alternates between rollout collection and parameter optimization. 
During rollouts, agents interact under an arrival-aware sampling scheme, locking completed agents to the STAY action to prevent invalid exploration. 
Upon collecting trajectories, the centralized Critic estimates advantages via GAE, and shared parameters $(\boldsymbol{\psi}, \boldsymbol{\phi})$ are updated via mini-batch SGD for $E_{\text{ppo}}$ epochs.
\begin{algorithm}[t]
\caption{MAPPO Training}
\label{alg:mappo_training}
\renewcommand{\algorithmicrequire}{\textbf{Input:}}
\renewcommand{\algorithmicensure}{\textbf{Output:}}
\begin{algorithmic}[1]
\REQUIRE Pre-trained GNN beamformer $f_{\theta}$, Actor--Critic networks $(\pi_\psi, V_\phi)$, total traning steps $T_{\text{total}}$, rollout length $T_{\text{rollout}}$, discount factor $\gamma$, GAE smoothing factor $\lambda$, PPO clipping factor $\epsilon$, update epoch number $E_{\text{ppo}}$, mini-batch size $B_{\text{ppo}}$;
\ENSURE Policy parameters $\psi$ and value parameters $\phi$;
\STATE Initialize shared-parameter Actor--Critic $(\pi_\psi, V_\phi)$;
\STATE Reset environment, obtain initial observations $\mathbf{o}[0]=\{o_n[0]\}_{n\in\mathcal{N}}$ and global state $s[0]$;
\FOR{$\text{step number}=1$ to $T_{\text{total}}$}
    \STATE Initialize rollout buffer $\mathcal{D}_{\text{ppo}}$;
    \FOR{$t=0$ to $T_{\text{rollout}}-1$}
        \FOR{each UAV $n \in \mathcal{N}$}
            \IF{UAV $n$ has arrived}
                \STATE $a_{n}[t] \leftarrow \text{STAY}$, $m_{n,t} \leftarrow 0$;
            \ELSE
                \STATE Sample $a_{n}[t] \sim \pi_{\psi}(\cdot \mid o_{n}[t])$, set $m_{n,t} \leftarrow 1$;
            \ENDIF
        \ENDFOR
        \STATE Execute joint action $\boldsymbol{a}[t]$, observe rewards $\mathbf{r}[t]$, termination $done[t]$, next observations $\mathbf{o}[t+1]$, and state $s[t+1]$;
        \STATE Store transition $\{\boldsymbol{o}_n[t], \boldsymbol{s}[t], a_n[t], r[t], done[t], $ \\ 
        $\log \pi_{\psi_{\text{old}}}(a_n[t]|\boldsymbol{o}_n[t]), V_{\phi_{\text{old}}}(\boldsymbol{s}[t]), m_{n,t}\}$ into $\mathcal{D}_{\text{ppo}}$;
    \ENDFOR

    \STATE Compute GAE advantages $\hat{A}_t$ and returns $\hat{R}_t$;

    \FOR{epoch $=1$ to $E_{\text{ppo}}$}
        \STATE Shuffle data in $\mathcal{D}_{\text{ppo}}$ and split into mini-batches;
        \FOR{each mini-batch $\mathcal{B}_{\text{ppo}} \in \mathcal{D}_{\text{ppo}}$}
            \STATE Compute $\mathcal{L}_{\text{actor}}(\psi)$, $\mathcal{L}_{\text{critic}}(\phi)$, and $\mathcal{L}_{\text{total}}(\psi, \phi)$;
            \STATE Update $(\psi,\phi)$ by minimizing $\mathcal{L}_{\text{total}}(\psi,\phi)$:
            \STATE $(\psi,\phi) \leftarrow (\psi,\phi) - \alpha \nabla_{\psi,\phi} \mathcal{L}_{\text{total}}(\psi,\phi)$;
        \ENDFOR
    \ENDFOR
\ENDFOR
\end{algorithmic}
\end{algorithm}

The trajectory planning subproblem's complexity is dominated by observation processing and Actor network forward propagation. 
Specifically, constructing the local observation $\boldsymbol{o}_n[t]$ requires aggregating coordinates from all $K$ users to calculate the relative average location $\Delta \boldsymbol{l}_{n,\text{avg}}^{\text{U}}[t]$, yielding an $\mathcal{O}(K)$ complexity. 
The Actor network then maps this observation to an action with a constant $\mathcal{O}(L_{a} d_{\text{actor}}^2)$ complexity. 
While the total computational complexity for $N$ UAVs is $\mathcal{O}(N(K + L_{a}d_{\text{actor}}^2))$, the CTDE framework enables parallel decentralized execution. 
Thus, the critical inference latency depends on the single-agent complexity $\mathcal{O}(K + L_{a}d_{\text{actor}}^2) \approx \mathcal{O}(K)$. 
Combined with the $\mathcal{O}(K)$ GNN beamforming complexity (Section \ref{gnn_complexity}), the overall per-slot decision latency scales linearly with $K$. 
This linear scalability offers a significant advantage over polynomial-time iterative algorithms, ensuring feasibility for real-time deployment in dynamic networks.

\section{Simulation Results}
\label{simulation}
\subsection{Simulation Settings}
\label{settings}
We consider a multi-UAV downlink communication network deployed over a $200 \times 200 \text{ m}^2$ square area, where ground users are randomly distributed and UAVs move between predefined start and destination locations. 
Each UAV is equipped with a ULA consisting of $L = 4$ antennas with half-wavelength spacing ($d_a = \lambda / 2$). 
The maximum transmit power of each UAV is $P_{\max} = 30 \text{ dBm}$, and the maximum communication coverage radius is set to $R_c = 150 \text{ m}$. 
The wireless channel follows the LoS channel model described in Section \ref{model}, with a reference channel gain of $\beta_0 = -20 \text{ dB}$ at $d_0 = 1 \text{ m}$ and effective receiver noise power of $\sigma_0^2 = -90 \text{ dBm}$.

The main hyperparameter are summarized in Table \ref{tab:sim_params}. 
These values follow commonly used settings in graph learning and PPO-based MARL and were selected to ensure stable convergence in our tested scenarios. 
For the inner-loop beamforming module, the GNN uses $L_{g}=3$ message-passing layers and is trained on $10000$ clustered channel samples using Adam for $E_{\text{gnn}}=25$ epochs to ensure convergence.
For the outer-loop trajectory module, MAPPO interacts with the environment for a $T_{\text{total}}=2 \times 10^{6}$ time steps with rollout length $T_{\text{rollout}}=256$.
During each policy update, the collected samples are divided into $4$ mini-batches and optimizad for $E_{\text{ppo}}=10$ epochs.

\begin{table}[t]
    \centering
    \renewcommand{\arraystretch}{1.3}
    
    \caption{Simulation Parameters}
    \label{tab:sim_params}
    \begin{tabular}{l|l}
        \hline
        \textbf{Parameter} & \textbf{Value} \\
        \hline
        GNN Embedding Dimension ($d_{\text{gnn}}$) & $128$ \\
        % Number of GNN Message-Passing Layers ($L_{g}$) & 3\\
        GNN Learning Rate ($\eta$) & $1 \times 10^{-3}$ \\
        GNN Weight Decay & $1 \times 10^{-6}$ \\
        MAPPO Learning Rate ($\alpha$) & $3\times10^{-4}$\\
        MAPPO Discount Factor ($\gamma$) & $0.99$ \\
        MAPPO GAE Smoothing Factor ($\lambda$) & $0.95$ \\
        MAPPO Clip Coefficient ($\epsilon$) & $0.2$ \\
        MAPPO Value Loss \& Entropy Coefficient ($c_1$ \& $c_2$) & $0.5$ \& $0.01$\\
        \hline
    \end{tabular}
\end{table}

\subsection{Beamforming Performance Comparison}
\label{GNN perform}
To evaluate the proposed GNN beamformer, we compare it with one optimization-based heuristic and two permutation-invariant learning baselines:
\begin{enumerate}
\item \textbf{Genetic Algorithm (GA)\cite{GA}:} A heuristic solving non-convexity via extensive space search. Implemented via PYMOO \cite{PYMOO} with 2,000 generations, which serves as a strong heuristic reference for small and medium-sized instances.
\item \textbf{DeepSets\cite{DeepSets}:} A neural network for set-structured data that ensures permutation invariance by encoding individual user features and aggregating them via global max-pooling.
\item \textbf{Multi-Scale PointNet (MS-PointNet)\cite{MS-PointNet}:} Adapted from point cloud processing, this convolutional network employs parallel 1-D convolutions with varying kernel sizes (e.g., 1, 3, 5) to capture multi-scale features while maintaining permutation invariance.
\end{enumerate}

To evaluate scalability and generalization, all learning-based models (Proposed GNN, DeepSets, and MS-PointNet) are trained on a network with $N=3$ UAVs and $K=12$ users. 
The post-training models are deployed to various test scenarios (e.g., varying UAV/user numbers) without re-training. 
This protocol evaluates whether the learned models capture transferable interference-management patterns rather than memorizing a single network size.

\subsubsection{Performance Comparison under Varying Channel Conditions}
\begin{figure}[t]
    \centering
    \includegraphics[width=0.45\textwidth]{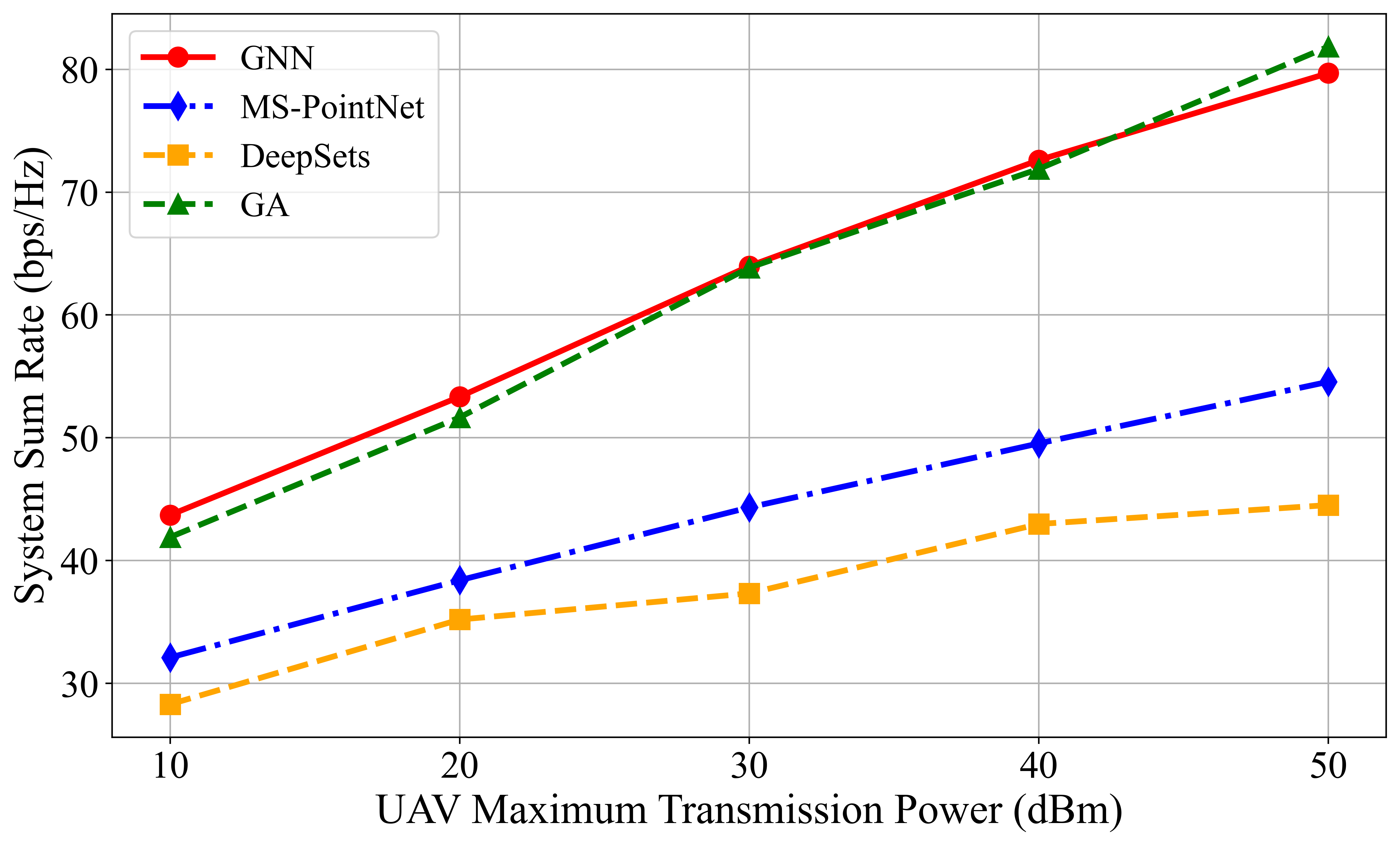}
    \caption{Sum rate vs. UAV maximum transmission power.}
    \label{fig:power_sweep}
\end{figure}
To further evaluate the proposed GNN beamformer's robustness, we examine the sum rate versus maximum transmit power $P_{\max}$ and noise power $\sigma_0^2$.
 
As shown in Fig. \ref{fig:power_sweep}, all methods benefit from higher transmit power, but the gain becomes architecture-dependent in the interference-limited regime.
A distinct performance divergence emerges in the high-power regime: the proposed GNN remains constantly close to GA ($\approx 80$ bps/Hz at $50$ dBm), whereas MS-PointNet and DeepSets saturate early at $55$ and $45$ bps/Hz, respectively. 
This behavior is consistent with the structural difference between the models: the baselines compress user-specific features into a fixed-size vector, losing the structural granularity required for precise interference nulling. 
In contrast, the proposed GNN leverages message passing to explicitly capture pairwise interference couplings, enabling effective suppression even under strong interference.

\begin{figure*}[t]
    \centering
    \begin{minipage}[t]{0.32\textwidth}
        \centering
        \includegraphics[width=\linewidth]{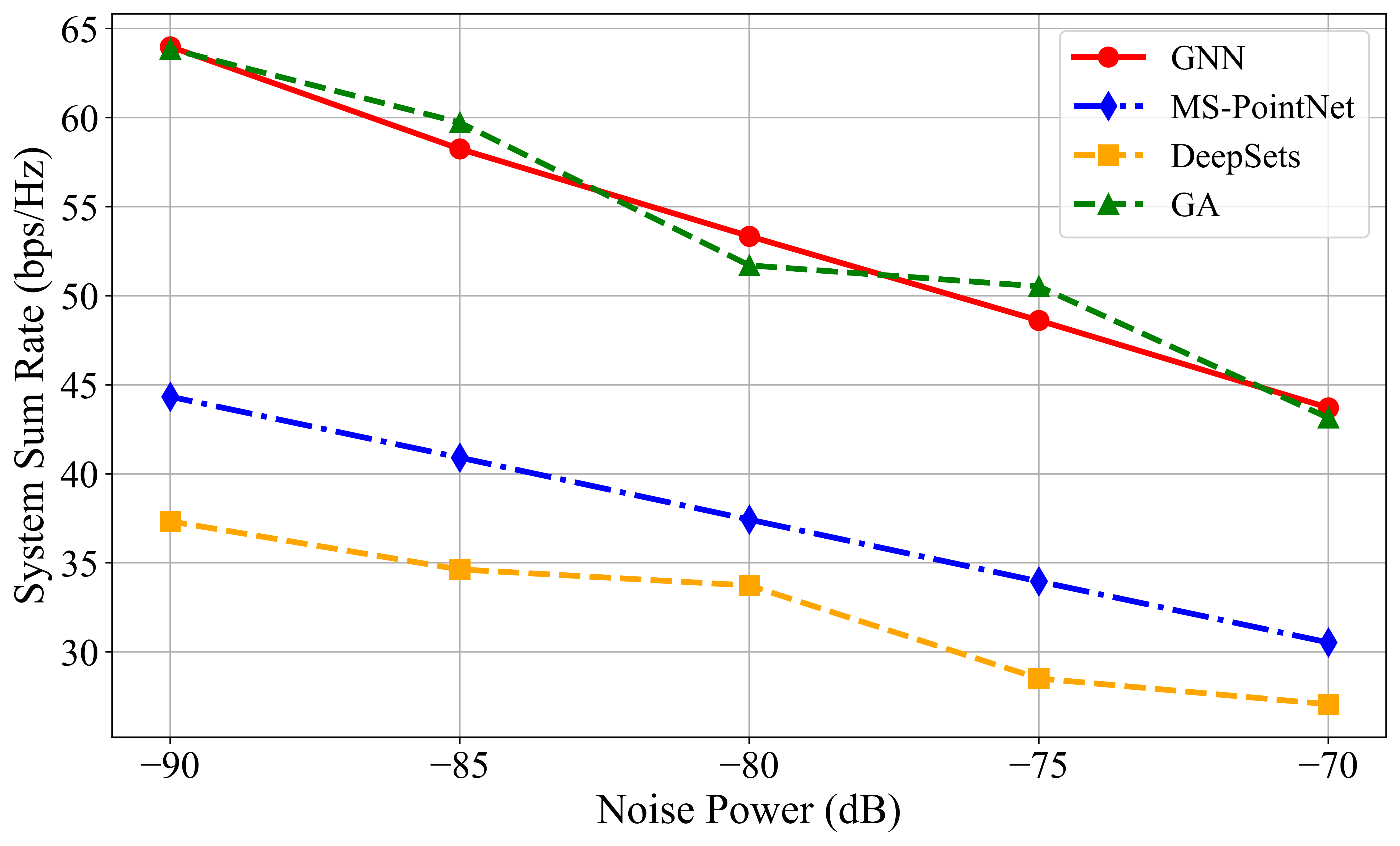}
        \caption{Sum rate vs. noise power.}
        \label{fig:noise_sweep}
    \end{minipage}
    \hfill 
    \begin{minipage}[t]{0.32\textwidth}
        \centering
        \includegraphics[width=\linewidth]{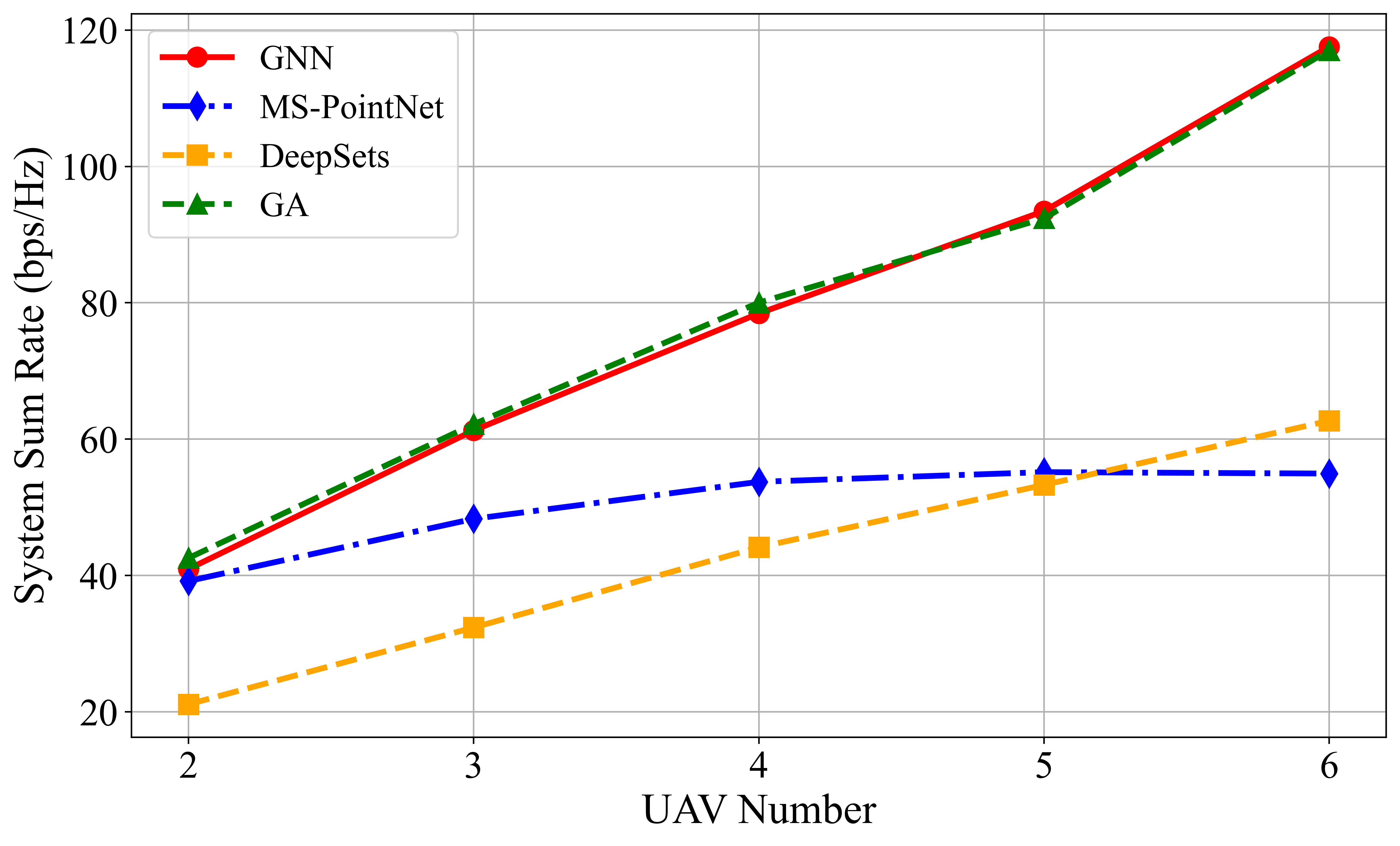}
        \caption{Sum rate vs. UAV number.}
        \label{fig:uav_gen}
    \end{minipage}
    \hfill 
    \begin{minipage}[t]{0.32\textwidth}
        \centering
        \includegraphics[width=\linewidth]{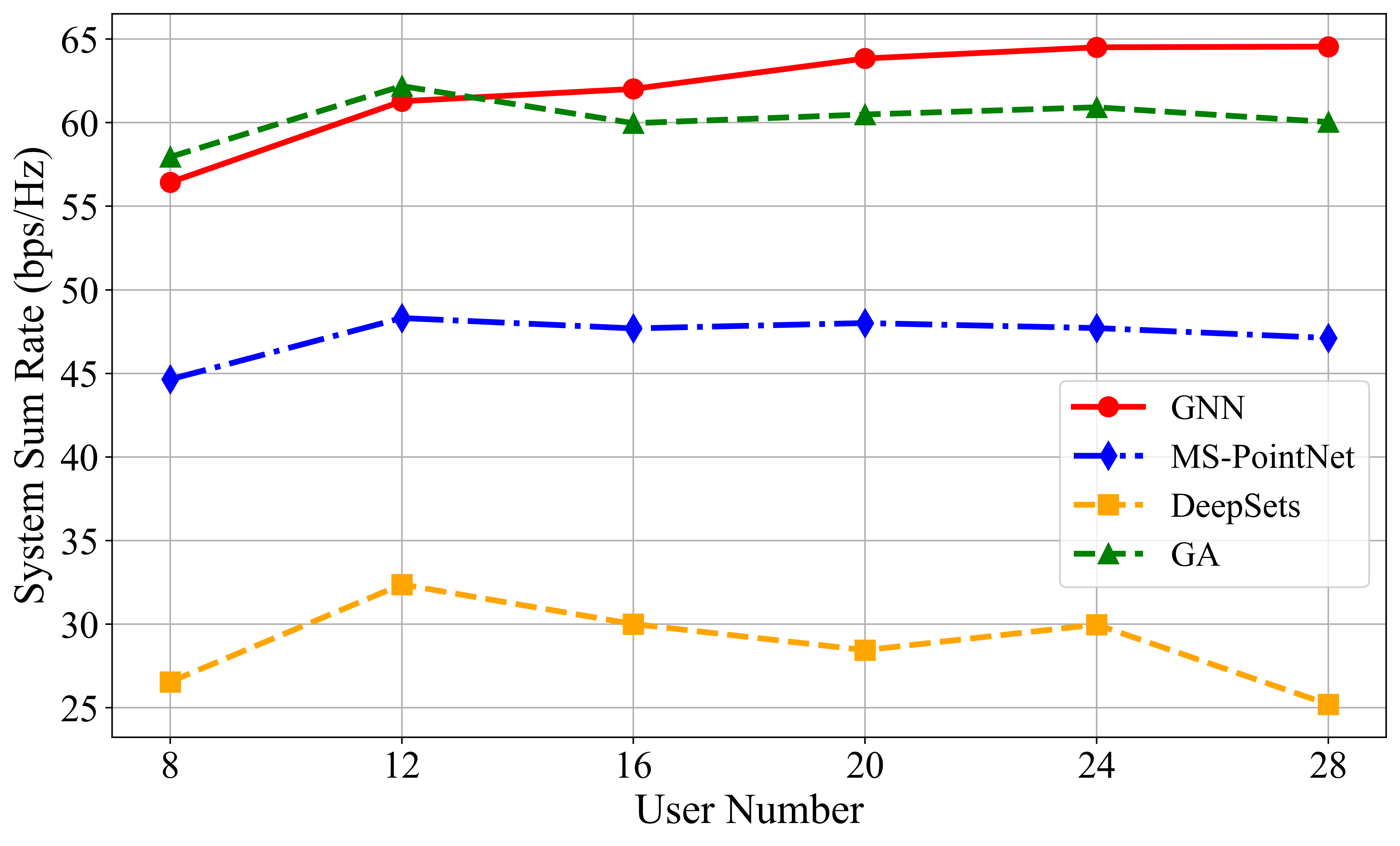}
        \caption{Sum rate vs. user number.}
        \label{fig:User_gen}
    \end{minipage}
\end{figure*}

Fig. \ref{fig:noise_sweep} depicts system robustness against noise power. 
The GNN demonstrates exceptional resilience, maintaining statistical parity with the GA bound even in high-noise settings ($-70$ dBm).
This indicates that the learned beamformer is robust not only to topology variation but also to moderate SINR degradation.
Conversely, the pooling-based baselines lag behind by a substantial margin of $15\text{--}20$ bps/Hz. 
This gap highlights the limitations of structure-agnostic architectures: without explicit topology modeling, they struggle to distinguish interference patterns from background noise. 

\subsubsection{Generalization Capability}
\label{Generalization}
To validate the generalization capability, we apply the pre-trained GNN, DeepSets, and MS-PointNet models directly to scenarios with varying UAV and user numbers.

Fig. \ref{fig:uav_gen} evaluates the system sum rate with respect to the number of the UAV. 
The proposed GNN demonstrates robust scability by preserving a small gap to the GA reference as the network becomes larger.
This indicated that the learned message-passing rule transfers well across different graph sizes.
In contrast, the performance of DeepSets and MS-PointNet degrades more noticeably as the number of UAVs increases, which reflect the difficulty of capturing multi-user interference using pooled set features.

Fig. \ref{fig:User_gen} illustrates the impact of the number of users and reveals a clear divergence in algorithmic behaviors. 
In particular, in the high-density regime ($K > 16$), the GNN outperforms the GA. 
Although this may appear counterintuitive, it reflects the curse of dimensionality in heuristic optimization.
As the search space grows exponentially with $K$, the interference landscape becomes increasingly coupled and highly non-convex.
Under practical computational constraints, global search methods such as GA are therefore more likely to converge permaturely to poor local optima.
In contrast, the GNN relies on explicitly learned local interference management rules, enabling superior inference in dense networks. 
Meanwhile, DeepSets shows an unusual downward trend, while MS-PointNet achieves only limited improvement.
Form a theoretical perspective, increasing users should yield multi-user diversity gains. 
However, DeepSets suffers from a severe information bottleneck: its global pooling operation compresses all $K$ user channels into a fixed-size latent vector. 
In dense networks, this aggressive compression blurs fine-grained spatial details, losing the precise information and causes the loss of precise angular details required to suppress closely spaced interference.  
Similarly, although MS-PointNet extracts local features, it lacks explicit edge-level message passing mechanisms to model the interactions between interfering and affected users. 
Consequently, severe intra-cluster interference from imprecise beamforming degrades overall performance.

\subsubsection{Computational Complexity Analysis}
\begin{figure}[t]
    \centering
    \includegraphics[width=0.45\textwidth]{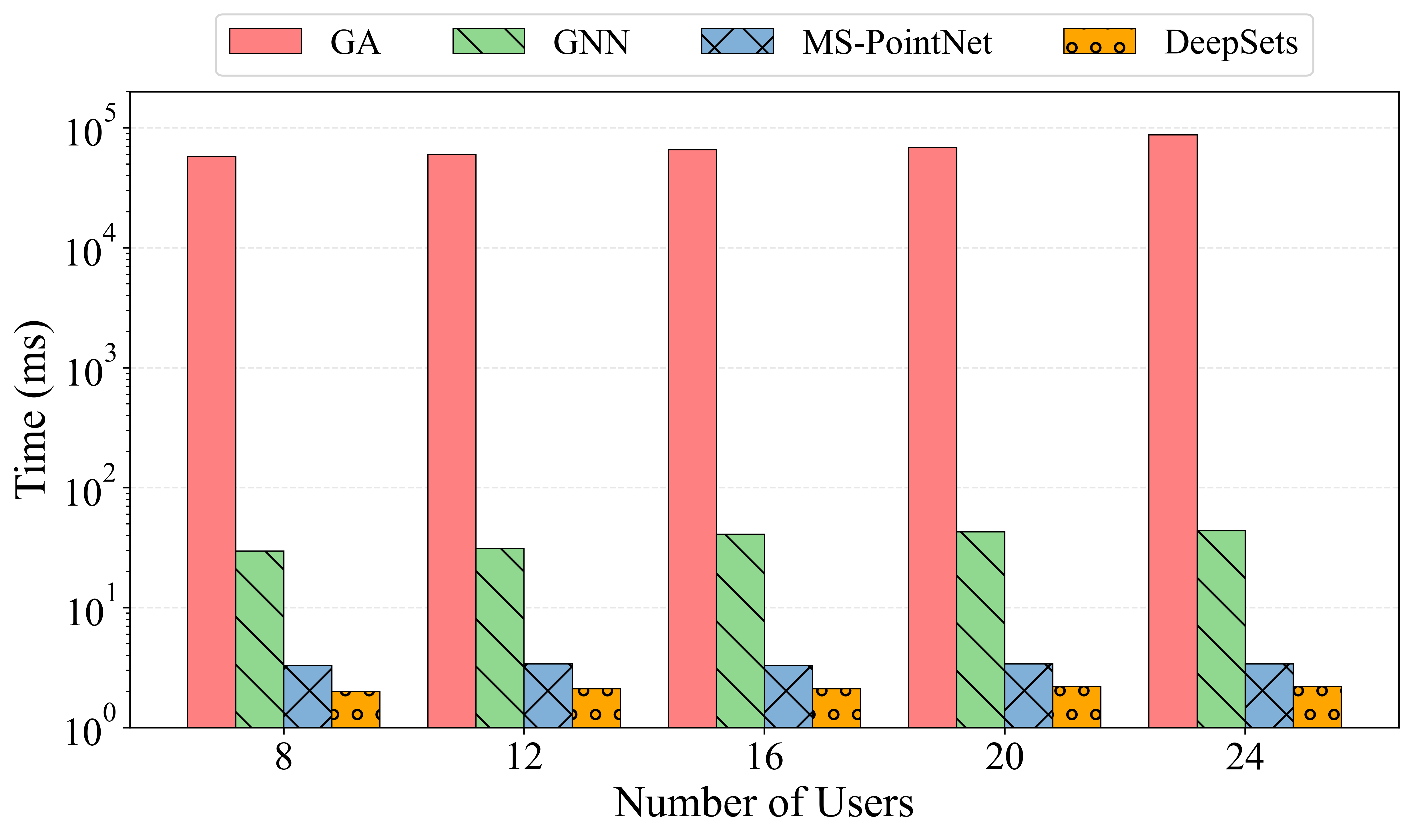}
    \caption{Average computation time per beamforming decision.}
    \label{fig:runtime}
\end{figure}
To assess real-time feasibility, Fig. \ref{fig:runtime} compares the average per-decision inference latency on a logarithmic scale. 
The GA requires $5.8 \times 10^4$ to $8.7 \times 10^4$ ms ($58$ to $87$ s) due to iterative population evolution. 
This prohibitive latency far exceeds high-mobility A2G channel coherence times, rendering GA impractical for online deployment where CSI expires within milliseconds. 
In contrast, the GNN achieves a stable inference time of $30$ to $45$ ms, representing a speedup of three orders of magnitude ($\approx 2000\times$) over GA. 
Crucially, the computational overhead exhibits excellent scalability with network density. 
As user numbers increase from $8$ to $24$, the GNN inference time grows only marginally from 29.6 ms to 43.8 ms. 
This modest linear increase confirms that graph-based inference remains computationally efficient even in dense networks. 
While DeepSets and MS-PointNet achieve lower latencies ($2$ to $4$ ms) via simplistic global pooling, they suffer severe performance degradation in dense networks as shown in Fig. \ref{fig:User_gen}. 
Since the GNN's latency remains well within the typical channel coherence window, it strikes the optimal balance between computational efficiency and beamforming accuracy.

\subsection{Trajectory Planning Performance Comparison}
To evaluate the MAPPO trajectory planner, the pre-trained GNN (Section \ref{GNN perform}) is fixed across all baselines to provide reward feedback, ensuring performance divergence stems solely from trajectory policies.
We simulate $3$ UAVs navigating from $\boldsymbol{l}_{\text{s}} = \{[20,20,100]^T, [20,180,100]^T, [180,180,100]^T\}$ to $\boldsymbol{l}_{\text{d}} = \{[180,100,100]^T, [180,20,100]^T, [20,100,100]^T\}$ with step size $D_{\text{fly}}=20$ m, safety distance $D_{\text{min}} = 25$ m, and horizon $T_{\max}=24$ steps. 
To evaluate the superiority of the proposed MAPPO framework, we compare it against five distinct baselines, including independent learning, value decomposition, and heuristic strategies:
\begin{enumerate}
    \item \textbf{Independent PPO (IPPO)\cite{MAPPO}:} A fully decentralized baseline where agents optimize policies based solely on local observations, serving to validate the necessity of centralized training.

    \item \textbf{Value-Decomposition Network (VDN)\cite{VDN}:} A value-based CTDE method that assumes the joint action-value function $Q_{\text{tot}}$ can be additively decomposed into local value functions, i.e., $Q_{\text{tot}} = \sum_{n=1}^K Q_n$.

    \item \textbf{QMIX\cite{QMIX}:} An advancement over VDN that employs a hypernetwork to mix local $Q$-values non-linearly, enforcing a monotonicity constraint ($\frac{\partial Q_{\text{tot}}}{\partial Q_n} \geq 0$) to ensure consistency between local greedy actions and the global optimum within the CTDE paradigm. 

    \item \textbf{Greedy Strategy:} A heuristic approach where UAVs exhaustively search the joint action space at each time step to maximize the instantaneous system sum rate.

    \item \textbf{Random Strategy:} A lower-bound baseline where UAVs select actions stochastically.
\end{enumerate}
To ensure a rigorous and fair comparison, all MARL-based algorithms (MAPPO, IPPO, VDN, and QMIX) share the modeling and weight settings detailed in Section \ref{settings}, differing only in learning paradigms. 
For heuristic baselines (Greedy and Random), a mandatory reachability constraint masks any action rendering the destination unreachable, guaranteeing all UAVs arrive within $T_{\max}$.

\begin{figure*}[t]
    \centering
    % 子图 (a)
    \begin{subfigure}[t]{0.32\textwidth}
        \centering
        \includegraphics[width=\linewidth]{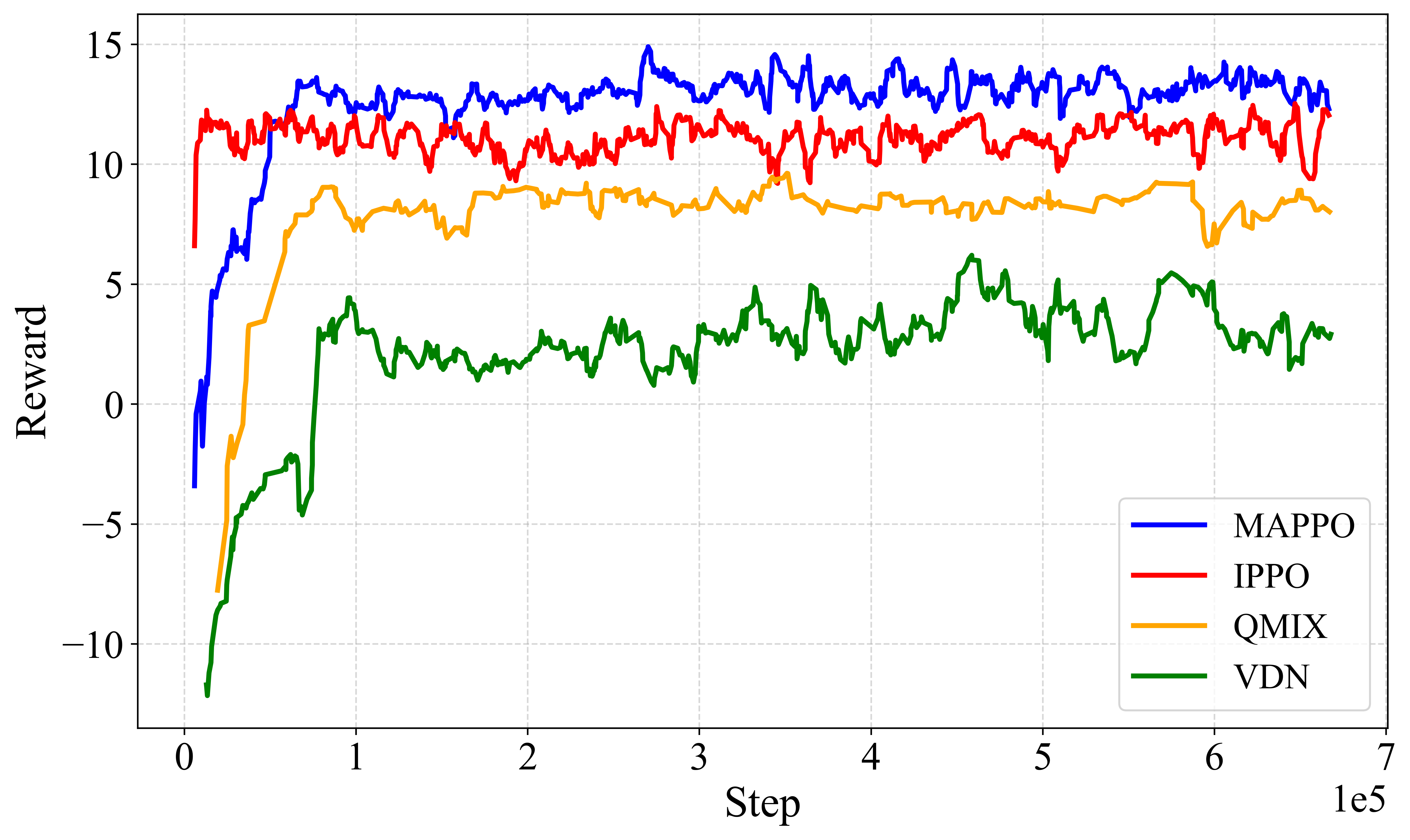}
        \caption{Convergence of average episode reward.}
        \label{fig:convergence}
    \end{subfigure}
    \hfill
    \begin{subfigure}[t]{0.32\textwidth}
        \centering
        \includegraphics[width=\linewidth]{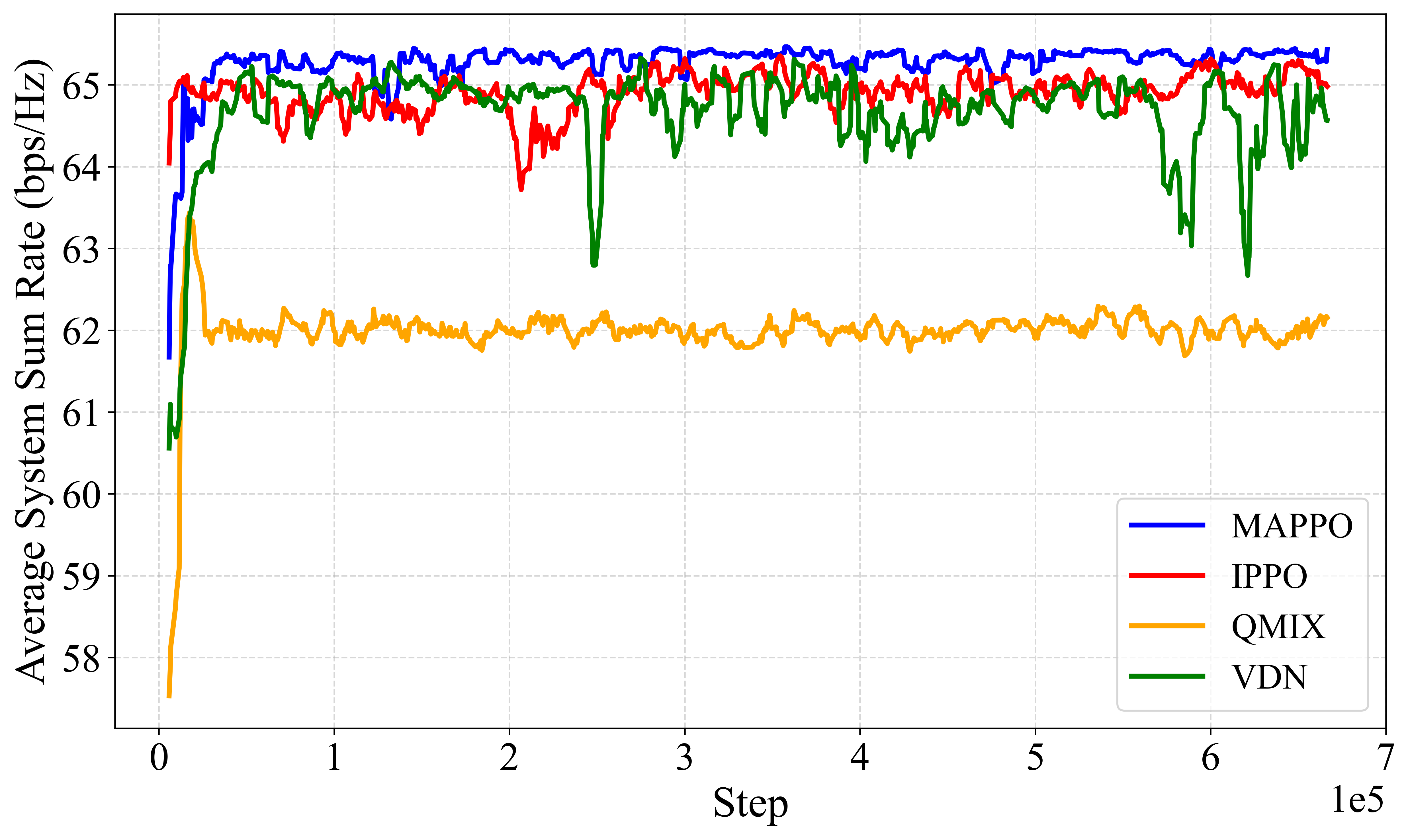}
        \caption{Evolution of average sum rate.}
        \label{fig:sum_rate_episode}
    \end{subfigure}
    \hfill 
    \begin{subfigure}[t]{0.32\textwidth}
        \centering
        \includegraphics[width=\linewidth]{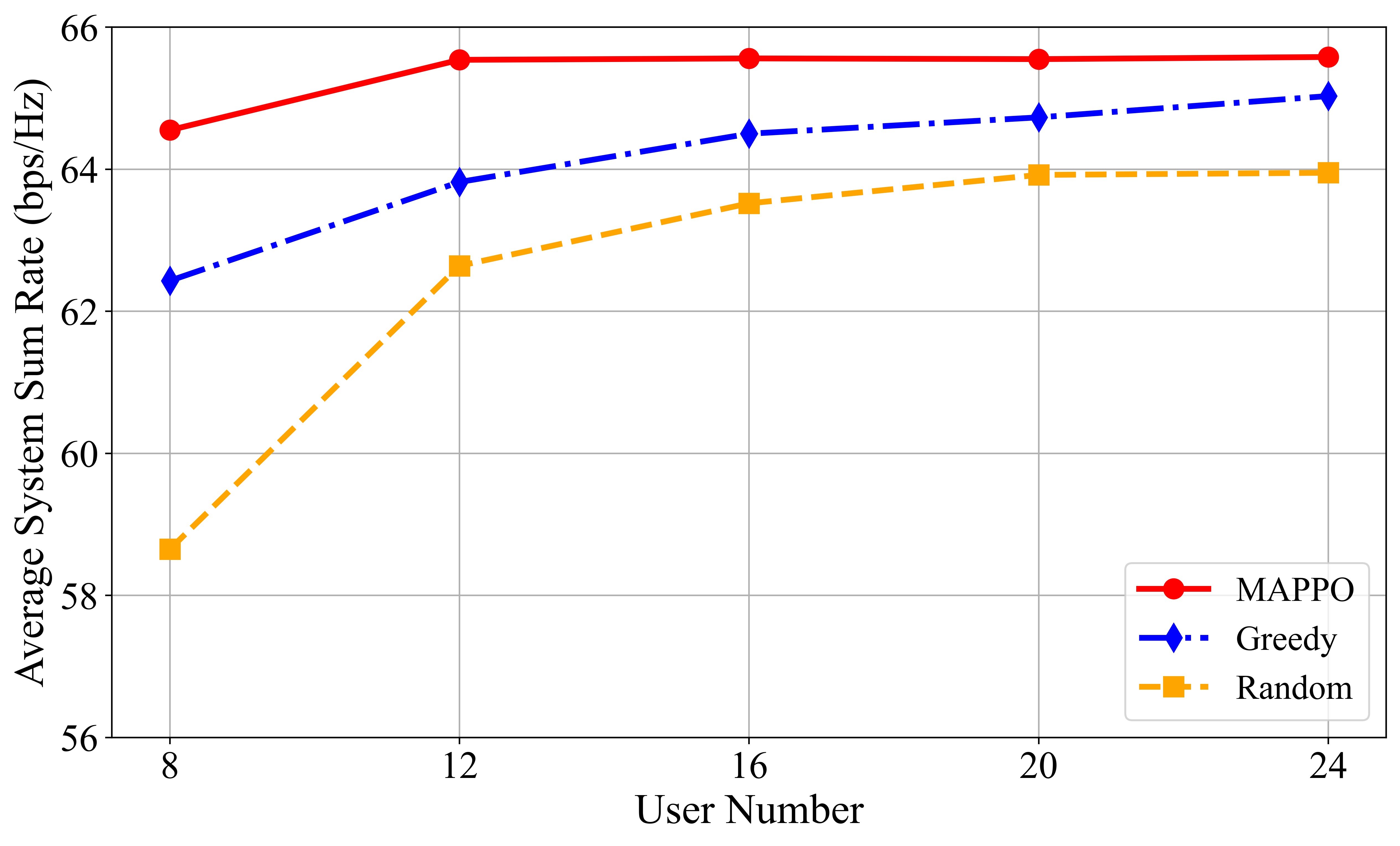}
        \caption{Average sum rate vs. user number.}
        \label{fig:heuristic_comp}
    \end{subfigure}

    \caption{Performance evaluation of the proposed MAPPO-based trajectory planning algorithm.}
    \label{fig:mappo_performance}
\end{figure*}

\subsubsection{Convergence Analysis}
Fig. \ref{fig:mappo_performance}(a) depicts the convergence curves of the average episode reward for the proposed MAPPO algorithm against three MARL benchmarks: IPPO, VDN, and QMIX. 
A detailed inspection of the learning dynamics reveals a distinct trade-off between initial speed and long-term optimality. 
IPPO exhibits a rapid initial surge, outperforming MAPPO within the first $5 \times 10^4$ steps. 
This stems from the myopic greedy nature of independent learners quickly learning local optimal actions. 
However, ignoring non-stationarity induced by other agents, IPPO fails to master cooperative evasion maneuvers, prematurely converging to local optima (rewards plateauing around $11$ to $12$). 
In contrast, MAPPO demonstrates superior long-term potential. 
Although the centralized critic initially requires more samples to accurately estimate the global value function, it eventually guides the UAVs to learn sophisticated cooperative strategies like collision avoidance. 
Consequently, MAPPO breaks independent learning performance bottlenecks, stabilizing at the highest reward interval ($13$ to $14$).
Furthermore, the results highlight the advantage of policy-based methods over value-decomposition approaches in this domain. 
VDN has the most unstable performance, suggesting that its linear value decomposition assumption is fundamentally insufficient to approximate the highly non-linear interference topology. 
While QMIX improves upon this via non-linear mixing, it still lags significantly behind the policy gradient-based algorithms. 
This indicates that for long-term trajectory planning problems involving complex coupled constraints, directly optimizing the policy is more effective than indirectly approximating the joint value function.

\subsubsection{Long-Term Communication Performance Analysis}
To assess the proposed framework's performance over long-duration missions, Fig. \ref{fig:mappo_performance}(b) tracks the long-term average system sum rate during the training phase. 
The results demonstrate MAPPO's clear dominance, rapidly ascending and stabilizing at a superior sum rate of approximately $65.5$ bps/Hz. 
This performance advantage stems fundamentally from the centralized training mechanism: by leveraging a critic to evaluate the global state, MAPPO effectively guides UAVs to learn sophisticated cooperative strategies, maximizing global throughput rather than myopic local gains. 
In contrast, IPPO, while occasionally reaching high peak rates, exhibits pronounced sawtooth-like oscillations. 
This instability reflects the limitations of  independent policy updates. 
Without global information sharing, agents fail to anticipate others' adaptive behaviors, creating a non-stationary learning environment where uncoordinated local improvements often destabilize the global objective. 
Furthermore, value-based decomposition methods prove less effective for this task. 
VDN suffers from severe performance fluctuations, indicating that its linear value decomposition assumption fails to approximate the highly nonlinear relationship. 
Similarly, while QMIX maintains a smoother learning curve, it converges prematurely to a suboptimal plateau of approximately $62$ bps/Hz. 
This stagnation suggests that the monotonicity constraint enforced by its mixing network restricts the model's expressiveness, preventing it from resolving the multi-constraint trajectory planning task.

\subsubsection{Performance Comparison with Heuristic Baselines}
Fig. \ref{fig:mappo_performance}(c) compares the long-term average system sum rate of MAPPO against two heuristic benchmarks (Greedy and Random) as the number of user increases from 8 to 24. 
The results indicate MAPPO consistently achieves the highest performance across all test densities. 
The Greedy strategy improves the instantaneous sum rate at each slot but remains myopic and cannot explicitly account for future positional advantages or mission completion constraints.
As a result, it may drive UAVs toward locally favorable regions that later become suboptimal. 
In contrast, MAPPO is trained to maximize the cumulative long-term return and can therefore trade off immediate rate improvement against future positional advantage. 
This enables the learned policy to favor trajectories that may be suboptimal in a single slot but beneficial over the full mission horizon.
This capability enables MAPPO to achieve global optimality over the entire flight. 
The Random policy performs worst, indicating that user-density growth alone is insufficient to deliver strong communication performance without purposeful mobility control. 
These findings validate the critical advantage of RL in solving long-horizon, continuous-space planning tasks where instantaneous greedy decisions prove suboptimal.

\subsubsection{Trajectory Visualization}
\begin{figure*}[t]
    \centering
    \subfloat[Sparse Scenario ($K=12$)]{\includegraphics[width=0.32\textwidth]{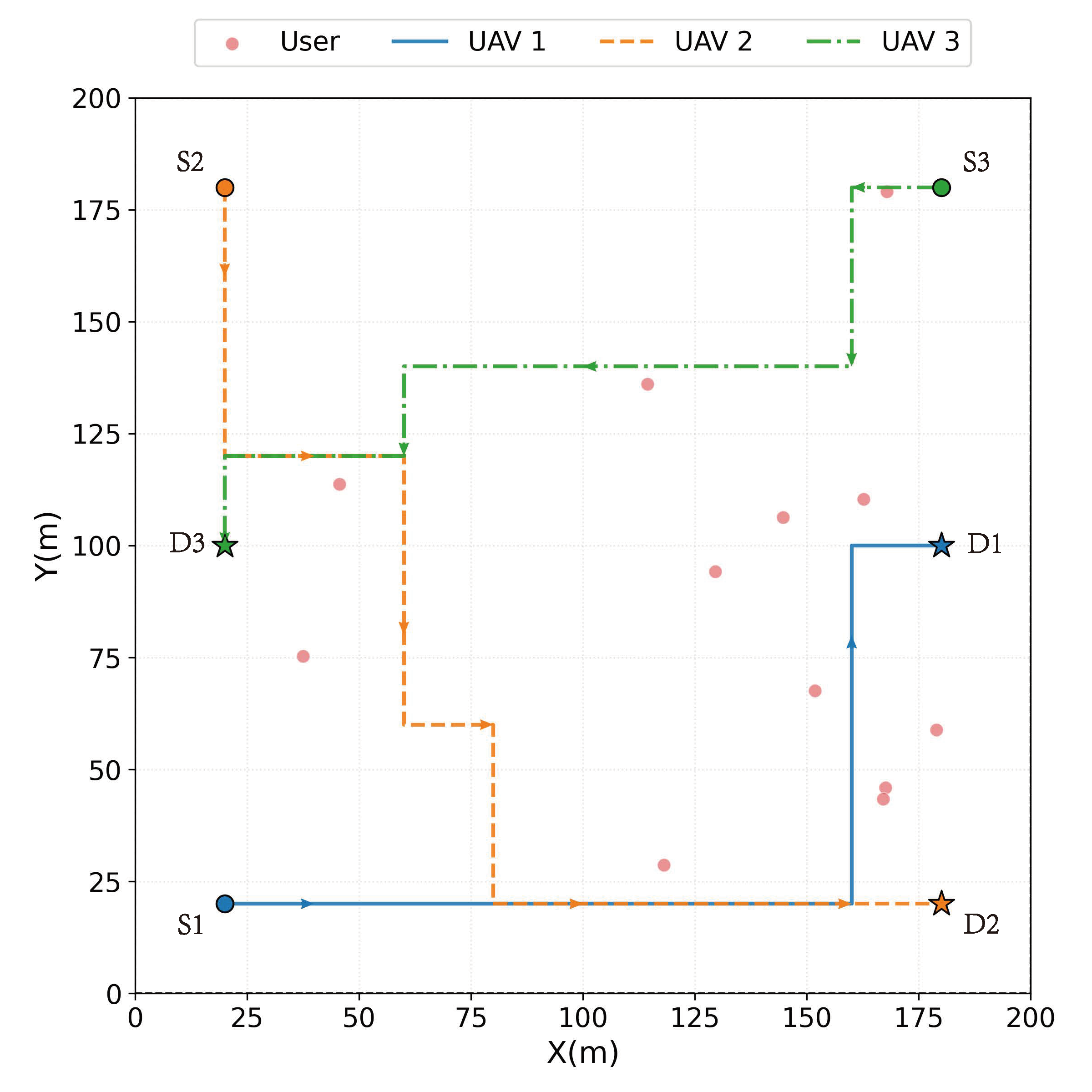}%
    \label{fig:traj_a}}
    \hfil
    \subfloat[Medium Scenario ($K=18$)]{\includegraphics[width=0.32\textwidth]{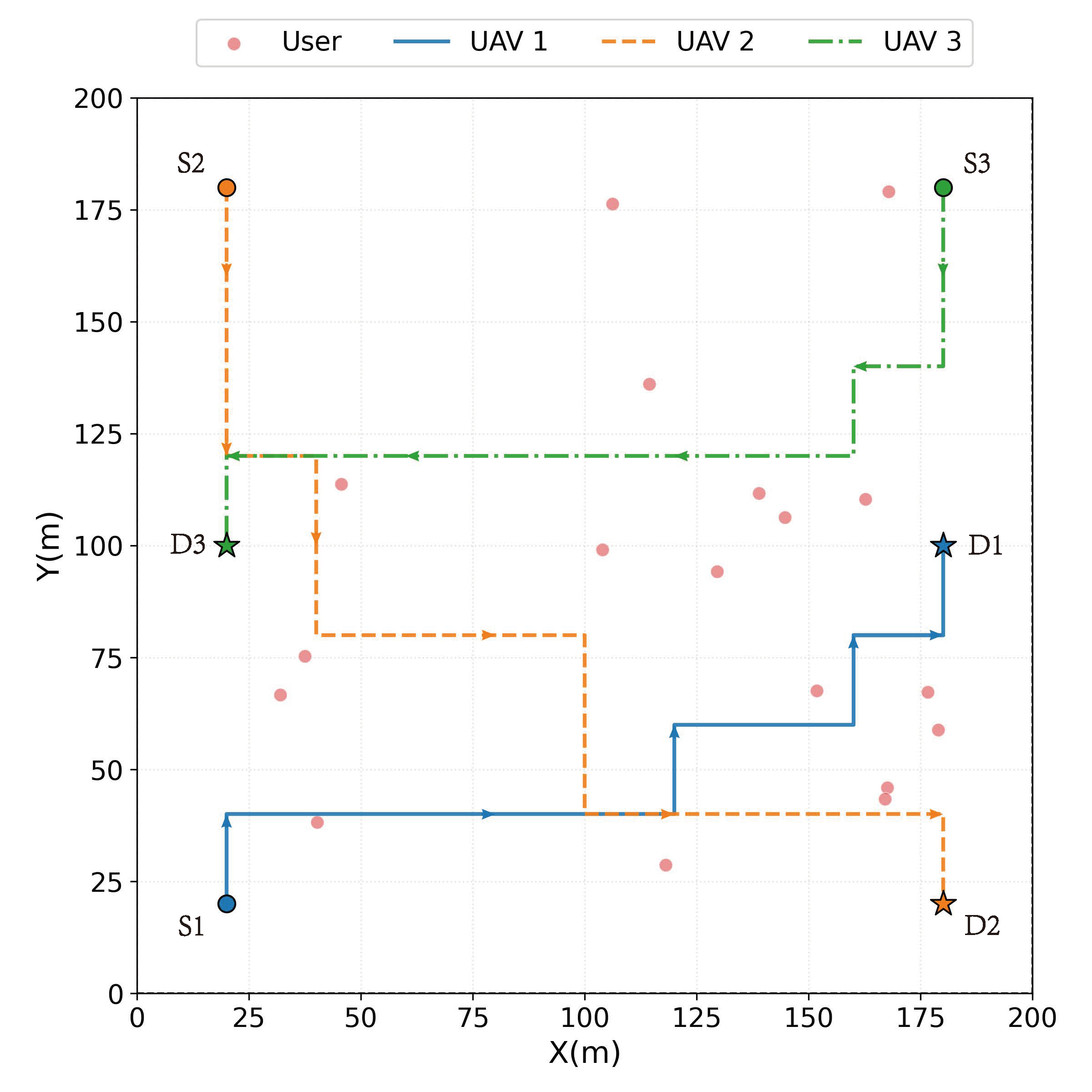}%
    \label{fig:traj_b}}
    \hfil
    \subfloat[Dense Scenario ($K=24$)]{\includegraphics[width=0.32\textwidth]{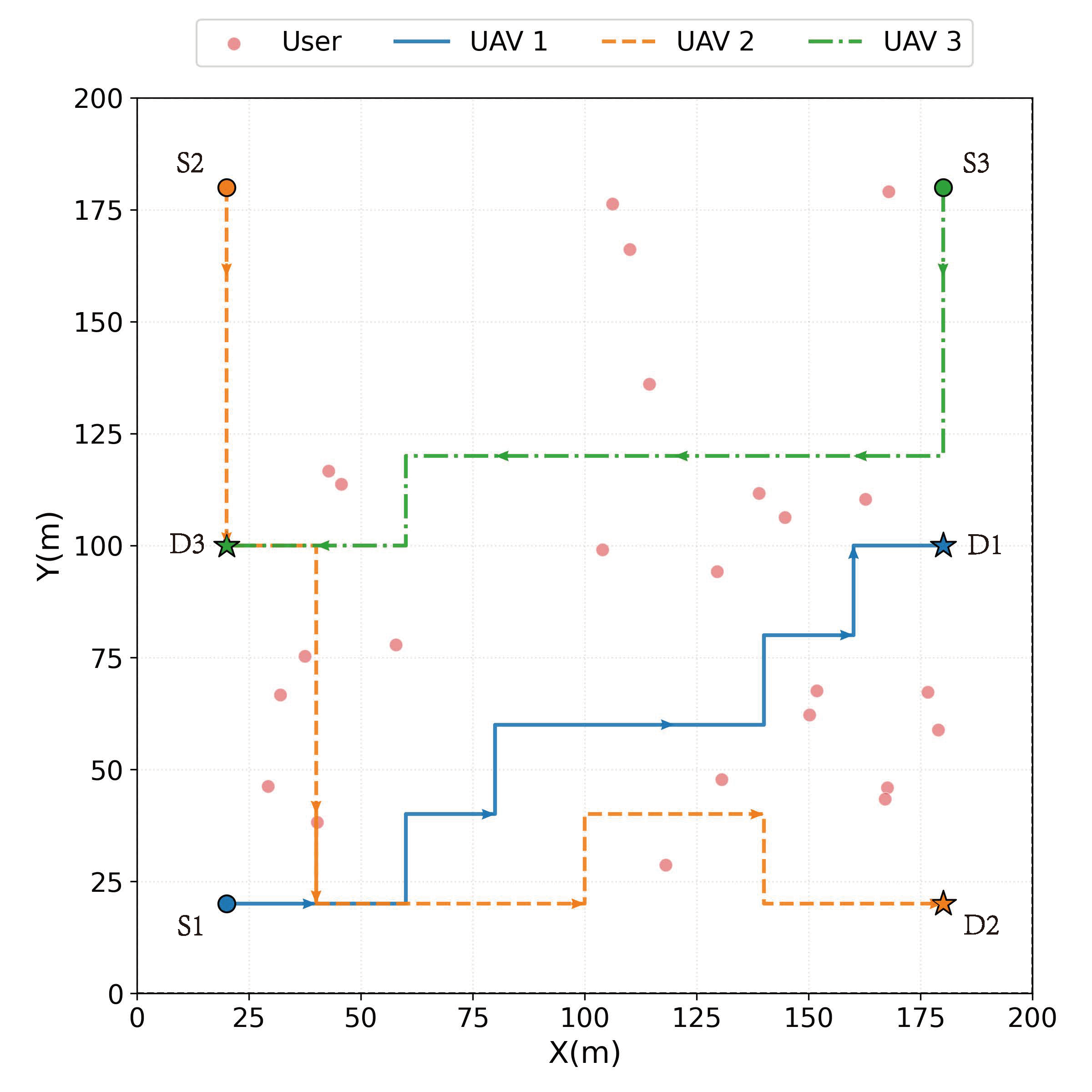}%
    \label{fig:traj_c}}
    \caption{Projected 2D trajectories of three UAVs navigating from start (S) to destination (D) under varying user densities.}
    \label{fig:trajectory_vis}
\end{figure*}

To visually uncover MAPPO's cooperative flight strategies and environmental adaptability, Fig. \ref{fig:trajectory_vis} visualizes the learned 2D UAV trajectories under sparse, medium, and dense user deployments.
Initially, observe that every UAV successfully reaches its destination within the deadline across all scenarios. 
The trajectories strictly adhere to boundary constraints without crossing, verifying robust collision avoidance learned via the feasibility-aware reward mechanism. 
Furthermore, flight paths maintain directional progress without redundant looping, minimizing total flight distance.
Beyond basic safety, the most distinct feature is service-aware maneuvering: UAVs deviate from the shortest straight-line paths. 
This intelligent, reward-driven deviation prompts UAVs to proactively detour towards user-dense regions, shortening service distances and establishing higher-quality LoS links. 
Crucially, this maneuvering evolves significantly with user density. 
In the sparse scenario (Fig. \ref{fig:trajectory_vis}(a)), trajectories are relatively linear, reflecting a distance-minimizing strategy since scattered users are adequately covered without significant deviation. 
However, in the medium scenario (Fig. \ref{fig:trajectory_vis}(b)) and dense scenario (Fig. \ref{fig:trajectory_vis}(c)), trajectories become increasingly intricate to maximize the sum rate.
Taking UAV 1 as a prime example: in Fig. \ref{fig:trajectory_vis}(a), it follows a near-optimal straight path along the bottom edge. 
In contrast, in Fig. \ref{fig:trajectory_vis}(c), it executes a substantial upward detour towards the center. 
This lateral maneuvering enables better coverage of dense user clusters in the bottom-middle region before proceeding to the destination. 
This confirms that MAPPO agents dynamically trade off flight distance for system sum rate maximization based on user distribution.

\section{conclusion and future work}
\label{cons}
In this paper, we proposed a hierarchically decoupled joint optimization framework for multi-UAV downlink communication networks, addressing the challenges of strong trajectory-beamforming coupling and dynamic network topologies. 
By exploiting the inherent timescale difference between fast channel variation and slower UAV mobility, the original problem was decomposed into two coordinated subproblems: instantaneous beamforming and long-term trajectory planning.
For the fast timescale, a topology-aware GNN beamformer was developed to capture the heterogeneous structure of UAV-user associations and intra-cluster interference, which enable scalable and low-latency beamforming decisions.
For the slow timescale, a CTDE-based MAPPO scheme was adopted to learn intelligent cooperative behaviors, such as service-aware maneuvering and implicit load balancing, to maximize long-term system throughput. 
Extensive simulation results demonstrated that the proposed framework significantly outperforms conventional optimization heuristics and deep learning baselines in terms of achievable sum rate, convergence behavior, and generalization across different network settings.
Future work will focus on extending the proposed framework to explicitly account for inter-cluster interference among multiple UAVs and to incorporate more realistic UAV mobility models with more complex motion dynamics.

\end{document}